# Condensation of Rocky Material in Astrophysical Environments

## Denton S. Ebel

Department of Earth and Planetary Sciences, American Museum of Natural History, New York, NY, 10024-5192
Department of Earth and Environmental Sciences, Lamont Doherty Earth Observatory of Columbia University, Palisades, NY, 10964-8000
Adjunct Graduate Faculty, Earth and Environmental Sciences, Graduate Center of City University of New York

**E-mail:** debel@amnh.org

-----------------------------------------------------------------------------------------------------
Volatility-dependent fractionation of the rock-forming elements at high temperatures is an early, widespread process during formation of the earliest solids in protoplanetary disks. Equilibrium condensation calculations allow prediction of the identities and compositions of mineral and liquid phases coexisting with gas under presumed bulk chemical, pressure and temperature conditions. A graphical survey of such results is presented for systems of solar and non-solar bulk composition. Chemical equilibrium was approached to varying degrees in the local regions where meteoritic chondrules, Ca-Al-rich inclusions, matrix and other components formed. Early, repeated vapor-solid cycling and homogenization, followed by hierarchical accretion in dust-rich regions, is hypothesized for meteoritic inclusions. Disequilibrium chemical effects appear to have been common at all temperatures, but increasingly so in less refractory meteoritic components. Work is needed to better model high-temperature solid solutions, indicators of these processes.

## 1. INTRODUCTION

It may be that much of the early solar nebula consisted of completely vaporized primordial material (> 1800 K) that cooled monotonically (*Cameron*, 1963; *Kurat*, 1988). This scenario explains the astonishing isotopic homogeneity (except O) of solar system materials from interplanetary dust grains (IDPs), to chondrites, to planets (e.g., *Zhu et al.*, 2001); but it appears to violate current theories about the thermal structure of the nebula and protoplanetary disk (*Cameron*, 1995; *Gail*, 1998; *Woolum and Cassen*, 1999). Yet equilibrium condensation sequences capture many first-order observations of the volatility-dependent fractionations of the elements, the identities of their host phases in meteorites, and even the chemical structure of the solar system (*Humayun and Cassen*, 2000). In the first edition of this volume, *MacPherson et al.*

(1988) concluded: "... good chemical and sparse textural evidence indicate that vapor-solid condensation played a major role in the genesis of many refractory inclusions in the early solar nebula. The condensation probably reflected neither perfect equilibrium nor perfect fractionation". Volatility-related fractionations among rocky elements were established at high temperatures, and are recorded in CAIs and chondrules (*Grossman*, 1996). The focus here is the vapor-liquid-solid equilibria relevant to these fractionation processes.

As investigations focus on disk structure, descriptions of chemical equilibrium between gas + silicate liquid + solid phases constrain the likelihood and extent of local processes that might occur in dense, cool disk regions subject to gravitational instabilities and shock heating (e.g., *Wood*, 1996; *Iida et al.*, 2001; *Desch and Connolly*, 2002; *Ciesla and Hood*, 2002); in hotter, partially ionized regions subject to magnetorotational turbulence and current sheet heating (e.g., *Joung et al.*, 2004); or in streams of material subject to solar flares and winds (the x-wind of *Shu et al.*, 2001). Parameterizations of chemical equilibrium calculations can be integrated into large-scale nebula models (e.g., *Cassen*, 2001). Condensation calculations are also applicable to other astrophysical environments, such as the stellar atmospheres that produce refractory interstellar dust (*Lattimer et al.*, 1978; *Lodders and Fegley*, 1997; *Ebel*, 2000). With new analytical techniques, meteoriticists have sharpened the search for pristine solar nebula condensates and their isotopic and trace element signatures, but also recognized that factors other than their thermal stability in a cooling gas have affected the chemistry of the most primitive objects found in meteorites.

The cosmochemical classification of the elements is based on their relative volatilities in a system of solar bulk composition. In a cooling (or heating) solar gas, elements are calculated to condense (or evaporate) in groups dependent upon their volatility: the abundant elements Ca, Al, and Ti are highly refractory; Si, Mg, and Fe are less refractory; K and Na are moderately volatile; S and C are more volatile. The rare earth (REE), platinum group (PGE) and other trace elements (e.g., highly volatile Hg, Tl) can be similarly classified (e.g., *Kerridge and Matthews*, 1988, part 7; *Ireland and Fegley*, 2000). These groups may be correlated, or combined, with *Goldschmidt's* (1954) geochemical classification of lithophile, siderophile, chalcophile, and atmophile elements (*Larimer*, 1988; *Hutchison*, 2004). The chemical compositions and associations of the most ancient minerals in meteorites strongly suggest that their formation was controlled in large part by the relative volatilities of the oxides of their constituent elements. Observed fractionations of the elements between meteoritic (asteroidal), cometary, and planetary reservoirs are also correlated to volatility (*Lewis* 1972a, 1972b). We model high-temperature equilibrium and disequilibrium processes to understand the formation of meteorite components, and the fractionation of the elements, one from another, in our solar system and in other stellar environments.

## 1.1. Chemical Equilibrium Calculations

"Condensation calculation" is a generic term for models describing the equilibrium distribution of the elements between coexisting phases (solids, liquid, vapor) in a closed chemical system with vapor always present. Predictions of chemical equilibrium made using the equations of state of the phases involved, for example at fixed total pressure ($P^{tot}$) and temperature (T), are independent of whether or not the system is cooling. They do not predict the sizes of resultant crystals. They can, however, be used to characterize the degree to which a chemical system is supersaturated in a potential condensate, which is a key parameter in understanding nucleation of condensates. Condensation calculations have been performed to describe chemistry in astrophysical environments since before digital computers became commonplace (*Latimer*, 1950; *Urey*, 1952; *Lord*, 1965; cf., historical review by *Lodders and Fegley*, 1997). *Lodders* (2003) has most recently computed the condensation temperatures of all the elements into pure, stoichiometric solid minerals from a cooling vapor of solar composition at $P^{tot} = 10^{-4}$ bar. Others have recently explored the stability of condensed liquid oxide solutions as well as solid solutions, all coexisting with vapor, in systems enriched in previously vaporized chondritic dust (e.g., *Yoneda and Grossman*, 1995; *Ebel and Grossman*, 2000). Chemical equilibrium calculations describe the identity and chemical composition of each chemical phase in the assemblage toward which a given closed system of elements will evolve, to minimize the chemical potential energy of the entire system. This knowledge is a prerequisite to any discussion of whether, where, how, and why equilibrium was or was not achieved in any particular scenario predicted by astrophysical models, or in relation to the origin of any particular meteoritic object.

Equilibrium calculations strictly apply only to systems that can be assumed to have reached or closely approached chemical equilibrium at a given point in their history. Disequilibrium effects can be explored by modification of equilibrium calculations, for example: fractionation or diffusion, by removing high-T condensates from further reaction at lower T (*Larimer*, 1967; *Larimer and Anders*, 1967; *Petaev and Wood*, 1998a,b; 2000 *Meibom et al.*, 2001); retarded nucleation, by constraining the onset of condensation for particular phases (*Blander et al.*, 2005; but cf. *Stolper and Paque*, 1986); evaporation, by computing saturation vapor pressures over evaporating phases, for input into kinetic models (e.g., *Grossman et al.*, 2000; *Nagahara and Ozawa*, 2000; *Alexander*, 2001; *Richter et al.*, 2002); photodissociation of gaseous molecules, by adjusting their thermal stabilities (*Ebel and Grossman*, 2001).

Reaction network calculations are a very different kind of simulation, in which the rates of individual reactions are addressed explicitly through rate constants. Parameters (e.g., reaction rates, sticking coefficients, surface energies of small clusters, heterogeneous catalytic effects of dust grains) for full gas-dynamical treatments (e.g., *Draine*, 1979; *Chigai et al.*, 2002) of heterogeneous reactions among rock-forming elements at high temperature are very poorly constrained. Network calculations appear to be most effective for low-temperature phase relations, and gas-phase reactions in the H-C-N-O-S system (e.g., *Hasegawa and Herbst*, 1993; *Bettens and Herbst*, 1995; *Gail*, 1998; *Aikawa et al.*, 2003; *Semenov et al.*, 2004). Partial ionization of gas would certainly affect condensation, and photodissociation would also occur in regions subject to ionizing radiation. Thermal ionization is not significant at the conditions discussed here, and calculations to the present do not address the condensation of rocky material in astrophysical environments with significant flux of ionizing radiation.

The formalism of classical thermodynamics is rooted in the phenomenology of materials, calibrated against more than a century of laboratory investigation (e.g., *Bowen*, 1928; *Nurse and Stutterheim*, 1950; *Newton*, 1987; *Ganguly*, 2001). Powers of interpolation, extrapolation, and prediction stem from this formalism (cf., *DeHeer*, 1986; *Anderson and Crerar*, 1993). This tool, and detailed observations of meteoritic inclusion and matrix chemistries and textures, constrain the conditions and processes these rocky materials might have experienced. Hypotheses for their formation must predict chemical, thermal and mechanical conditions that are in accord with those constraints. Pressures, temperatures, oxygen fugacities, and chemical environments are deduced from analyses of meteoritic materials and constrain astrophysical models for the protoplanetary disk in which those materials formed. The resulting understanding of the universal process of disk formation is central to learning how and where habitable planets form, and how unique our own Earth really is.

## 1.2.  Earlier Reviews and Present Focus

Several excellent reviews cover condensation of crystalline solid phases from a gas of solar composition (e.g., *Larimer*, 1988; *Lewis*, 1997; *Fegley*, 1999), particularly *Grossman and Larimer* (1974). *Davis* (2003) has compared kinetically controlled processes with equilibrium processes. The present chapter briefly describes data and algorithms for modeling the thermal stability of condensates relative to vapors of various compositions, presenting results in the form of phase diagrams with a discussion of their use. I also briefly discuss the meteoritic evidence that may indicate direct condensation, or the establishment of chemical equilibrium between gas, solid and liquid phases in the protoplanetary disk before accretion of the meteorite parent bodies.

Dust-gas chemistry has received substantial attention from astrophysicists interested in interstellar and cometary material (e.g., *Millar and Williams*, 1993; *Altwegg et al.*, 1999). The role of interstellar processes in establishing the meteorite record is increasingly recognized (e.g., *Wood*, 1998; *Irvine et al.*, 2000). Some grains found in meteorites were formed at high temperatures around stars with C/O ≥ 1, or in supernovae (cf., *Zinner*, 1998; *Nittler and Dauphas*, 2005). Astrophysical observations (e.g., *Chen et al.*, 2003) are yielding new data about stars of non-solar metallicity (i.e., the abundance of hydrogen relative to all the heavier elements; in astrophysics: *metals*. Here, *metal* will refer to Fe, Ni, Co, etc. and their alloys.). Measurement of the atmospheric compositions of stellar grain sources is an active field, as is modeling of nucleosynthesis in AGB stars and novae. Condensation calculations using these results as inputs are vital to understanding the origins of interstellar and presolar grains (e.g., *Amari et al.*, 2001; *Ebel and Grossman*, 2001; cf., review by *Lodders and Fegley,* 1997). Grain formation in stellar outflows is also central to understanding the volatility-dependent depletions of rock-forming elements observed in the interstellar medium (*Field*, 1974; *Irvine et al.*, 1987; *Ebel*, 2000; *Kemper*, 2002). Condensation calculations have been pursued down to absolute zero (*Lewis*, 1972a), however, the equilibrium model is of very limited utility at such low temperatures. High-temperature disequilibrium effects, such as temperature differences between gas and solids (e.g., *Arrhenius and Alfvén*, 1971), can be addressed by modification of equilibrium calculations. This chapter is about the fates of condensable atoms in the nebula and disk, at temperatures above 1000 K, and total pressures between $10^{-8}$ and 1.0 bar (number densities of molecules > $3 \times 10^{16}$ cm$^{-1}$). These are conditions, plausible in some but not in all astrophysical models of portions of the protoplanetary disk, where ensembles of atoms may be assumed to approach chemical equilibrium on timescales that make such calculations relevant and useful.

## 2.  TECHNIQUES

## 2.1.  Elemental Abundances

*2.1.1.  Solar and C-enriched Systems.*  The calculations here consider closed chemical systems containing the 22 elements having abundance greater than Zn, plus F: H, He, C, N, O, F, Ne, Na, Mg, Al, Si, P, S, Cl, Ar, K, Ca, Ti, Cr, Mn, Fe, Co, Ni. Equilibrium among these elements fixes their partial pressures in the vapors, from which thermal stabilities of trace element-bearing compounds can be calculated. The relative atomic fractions of the elements in a vapor of solar composition are from *Anders and Grevesse* (1989), which give a C/O ratio of 0.43. Recent work by *Allende Prieto et al.* (2002) resets solar C/O to 0.50, and *Lodders* (2003) has most recently re-evaluated solar abundances of all the elements (cf., *Grevesse and Sauval*, 1998). These revisions have very minor effects on the relative thermal stabilities of rocky condensates. Because C and O combine to form the highly stable CO gaseous molecule, the excess O makes a gas of solar composition oxidizing. If C/O > ~1.0 by addition of C to, or removal of $H_2O$ from, a gas of otherwise solar composition, the system becomes highly reducing, and very different solid assemblages become stable. C-enriched gas compositions are relevant to the formation of presolar graphite, SiC, TiC, and other grains in C-rich stellar environments (*Sharp and Wasserburg*, 1995; *Lodders and Fegley*, 1997), and possibly also to the formation of enstatite chondrites (*Ebel and Alexander*, 2005).

***2.1.2. Dust-enriched Systems.*** Innumerable scenarios can be explored for adjusting elemental abundances in a gas of solar composition, by the addition or subtraction of plausibly fractionated components. *Wood* (1963, 1967) proposed enrichment of vapor of solar composition by a fractionated dust of previously condensed rocky material, perhaps in the midplane of a protoplanetary disk to as far as 10 AU (*Millar et al.* 2003). In this scenario, dust consisting of oxides and sulfides of rock-forming elements is concentrated relative to $H_2$ and other volatile species. The resulting bulk composition is assumed to be heated to vaporization, and the condensation history of the cooling vapor is predicted. *Wood and Hashimoto* (1993) explored the condensation of bulk compositions enriched or depleted in refractory, organic, and icy components and in hydrogen. *Yoneda and Grossman* (1995) explored condensation in solar bulk compositions enriched by a dust of ordinary chondrite composition. Their objective was to find the conditions where CaO-MgO-Al$_2$O$_3$-SiO$_2$ (CMAS) liquids are stable. *Ebel and Grossman* (2000) extended this work to a vapor of solar composition, enriched in a dust of carbonaceous chondrite (Orgueil, CI) composition, and they calculated the thermal stability of CMAS-TiO$_2$-FeO-Na$_2$O-K$_2$O liquids and FeO-bearing silicates. The portion of a vapor of solar composition that *Ebel and Grossman* (2000) considered to be condensable as chondritic (CI) dust is $5.66 \times 10^{-4}$ (atomic), or $6.69 \times 10^{-3}$ (by mass), of a canonical solar vapor (*Anders and Grevesse*, 1989). Their dust enrichment factor of 1000xCI corresponds to a dust/gas mass ratio of ~6.7, which is not unjustifiable on astrophysical grounds. Chondritic material is highly depleted in C and H relative to solar. The addition of such "dust" to a gas of solar composition increases the partial pressure of oxygen in particular, but also condensable elements present in the dust.

## 2.2. Thermodynamic Data

If everyone used the same data and algorithms, there would be a unique canonical 'condensation sequence'. Such is not the case, and in recent years differences have been exaggerated by the employment of ever more complex models for mineral solid solutions: minerals where multiple combinations of elements can occupy sites in the same crystalline lattice. There is no single '*most* correct' data base, or set of solid solution models, however, there are carefully selected, internally consistent sets of data, and haphazardly selected sets collected from disparate sources. The former should be *more* correct, in application, than the latter. The power and limitations of a particular calculation are most clear when algorithms, but most importantly the core data, are described as completely as editors will allow. For example, the mixing of endmember olivine species fayalite (Fe$_2$SiO$_4$) and forsterite (Mg$_2$SiO$_4$) can be treated as a mechanical mixture of separately condensing endmembers, or as a real solid solution with accounting (based on experimental data) for the energetic consequences of ionic mixing on the olivine crystalline lattice. The particulars of data, and how it is implemented in calculation, are crucial to evaluating the results of such calculations.

There is reasonable agreement as to the first order phenomena predicted during equilibrium cooling of a gas of solar composition at fixed $P^{tot}$ (*Larimer*, 1967; *Grossman*, 1972; *Grossman and Larimer,* 1974; *Wood and Hashimoto*, 1993; *Yoneda and Grossman*, 1995; *Ebel and Grossman*, 2000; *Lodders*, 2003; see Section 3.1.2). Calculations of silicate liquid stability as a function of $P^{tot}$ and enrichment in chondritic dust also agree where models overlap (*Yoneda and Grossman*, 1995; *Ebel and Grossman*, 2000; *Ebel*, 2005).

An important consideration is the fundamental thermodynamic data for the elements, depending upon what forms of equations of state are used for solids and liquids. The equations of state for compounds used here combine their enthalpy and entropy of formation from the elements in their standard states at 298 K and 1 bar with the heat capacity of the compound at all relevant temperatures. From these data, suitable equations of state (e.g., apparent Gibbs energy of formation from the elements at constant P and T) are readily calculated from free energy (Giaque) functions for the constituent elements (cf., *Ghiorso and Kelemen*, 1987; *Berman*, 1988; *Anderson and Crerar*, 1993). Direct use of tabulated Gibbs energies (e.g., *Chase*, 1998) can be hazardous (cf., *Lodders*, 2004).

*2.2.1. Major Elements.* Equilibria between gas, solid and liquid phases are most conveniently calculated by referring all formation reaction energies to the monatomic gaseous species of the elements. The elements can be divided into major, minor, and trace categories based upon their absolute abundances. The abundant elements are atmophile (gaseous) H, C, N, O; lithophile ("rock-loving") Si, Mg, Al, Ca, Ni, Na, K, Ti; and siderophile ("iron-loving") Fe, Ni, Co, Cr, and S. Just the oxides of Si and Mg, with metallic Fe, constitute >85% of matter condensable above 400 K from a vapor of solar composition. Most of the other elements have minor effects on the identity and compositions of solids and liquids into which they substitute in equilibrium with vapor under most assumed astrophysical conditions.

*2.2.2. Minor and Trace Elements.* These elements usually occur as trace substituents in minerals, so they are frequently considered separately from the major, or abundant, rock-forming elements that make up those minerals. The rare earth elements (REE), U, Th, Li, and others are lithophile; siderophile trace elements include the platinum group elements (PGE: Ru, Rh, Pd, Re, Os, Ir, Pt). The moderately volatile elements (e.g., Mn, Cu, Na, Se, Zn, Cd) were defined by *Larimer* (1988) as "never depleted by more than a factor of 5 in any chondrite, relative to CI", and displaying "no detectable variations within a chondrite group" (cf., *Xiao and Lipschutz*, 1992). *Hutchison* (2004) delineates the 50% condensation T of Cr as separating refractory from moderately volatile (Au, Mn, alkalies) elements, and the onset of FeS condensation as marking the T below which 'highly volatile' elements (Pb, Tl, Bi, Cd, Hg, In) condense. In most calculations, these trace elements are allowed to condense as pure elemental, oxide, or sulfide solids, which are then assumed to dissolve into major element phases present at corresponding conditions. For example, most Cd condenses as CdS, at temperatures where FeS is stable (e.g., *Lodders*, 2003), hence CdS is assumed to dissolve into FeS. Equations of state for the gaseous and solid oxides and other compounds of many minor elements are poorly known or not readily available (*Davis and Grossman*, 1979). The relative solubilities of these elements between gas, solid and liquid phases are in most cases also not well understood, so analysis of meteoritic minerals usually guides the choice of host phase. Thus in "oxidized" systems (e.g., solar), the REE condense into perovskite or hibonite (*Lodders*, 2003), and in reduced systems into oldhamite (*Lodders and Fegley*, 1993; formulae in Table 1), based on measurements of those phases in ordinary and enstatite chondrites (e.g., *Crozaz and Lundberg*, 1995), and on the condensation temperatures of pure REE oxides and sulfides. The refractory siderophile elements including PGE are expected to condense into refractory metal nuggets at high temperatures (*El Goresy et al.*, 1979; *Fegley and Palme*, 1985; *Lodders*, 2003; *Campbell et al.*, 2001), and into Fe-Ni alloy when iron condenses. The behavior of trace elements becomes less predictable as their volatility increases, because the assumption of equilibrium becomes less tenable at lower T (*Kornacki and Fegley*, 1986).

**2.2.3. Gaseous Species.** The primary reservoir of the nebula and disk is a vapor containing hundreds of gaseous molecular species (e.g., H, C, O, $H_2$, CO, $CO_2$, $H_2O$). At and below $P^{tot}$ = 1 bar the gas can be considered as an ideal mixture of ideal gases. Because most gas species data is derived from vibrational spectroscopic measurements (*Chase*, 1998), it is assumed to be internally consistent. That is, an equilibrium calculation using data for reacting species accurately reproduces equilibria involving those species as observed in the laboratory. Algorithms to minimize the chemical potential energy in a large, speciated gaseous system stem from the work of *White et al.* (1958; cf., *Smith and Missen*, 1982).

**2.2.4. Pure Solid Species.** Of the 275 mineral species identified in meteorites (*Rubin*, 1997), only a small subset is abundant, stable at high-T, and has well-known thermodynamic properties. Much of the high-quality data for solid phases of geologic interest was determined many years ago, in labor-intensive systematic investigations by dedicated experimentalists. Mineral phases common in terrestrial geology are well-studied, however, data are absent for some high-temperature phases found in chondrites (e.g., rhönite; *Fuchs*, 1978; see *Beckett et al.*, 2005). Data used in calculations must account for phase transitions that occur in many solids (e.g., α-β quartz - cristobalite for $SiO_2$), as their most stable crystalline lattice structure changes with temperature. Most magnetic transitions are below the temperatures at which equilibrium calculations might be assumed to apply, and they are not considered here. Multiple data compilations and laboratory studies contain equations of state for mineral phases of interest (cf., *Berman*, 1988; *Kuzmenko et al.*, 1997). For example, for six published values for the enthalpy of formation of grossite from the elements at 298 K, the average is -4012.261 kilojoules, with a standard deviation of 11.413 kJ (*Berman*, 1983; *Geiger et al.*, 1988, and references therein). A similar selection exists for hibonite (*Kumar and Kay*, 1985) and perovskite (formulae in Table 1). These differences affect the stabilities of Ca-aluminates relative to competing oxides, and account for many differences in published condensation sequences (see Section 3.1.2), as reviewed in detail by *Ebel and Grossman* (2000). In nature, Mg and Ti dissolve into hibonite (*Hinton et al.*, 1988; *Beckett and Stolper*, 1994; *Simon et al.*, 1997; *Beckett et al.*, 2005), and this behavior is not captured in any calculations to date. Treating these phases as solid solutions is a goal of future work. Various permutations of all available data allow a variety of predictions for the relative stabilities of end-member (pure) corundum, hibonite, Ca-aluminate, grossite, spinel, and perovskite (Table 1) without appreciably altering the relative proportions of Ca, Al, and Ti oxides which are condensed at a particular T and $P^{tot}$. This is because the differences in the free energies for the solid phases are small, compared to the differences in free energies of the solids relative to the vapor.

**2.2.5. Solid Solutions.** Many minerals found in meteorites, and predicted to be thermodynamically stable at high temperatures in a gas of solar composition, are solutions of two or more end-member components with the same crystal structure. Because these solid solution minerals incorporate variable amounts of several elements depending upon the conditions in which they form, they are potentially powerful indicator minerals for constraining the condensation, evaporation and crystallization histories of meteoritic inclusions in which they occur (e.g., *Fraser and Rammensee*, 1982). Examples include:

melilite:   $Ca_2Al_2SiO_7$ - $Ca_2MgSi_2O_7$                    (gehlenite - åkermanite)
            substitution: $Al_2$ for MgSi
olivine:    $Mg_2SiO_4$ - $CaMgSiO_4$ - $Fe_2SiO_4$     (forsterite - monticellite - fayalite)
            substitution: Fe and Ca for Mg
metal alloy:  Fe,Cr,Co,Si,Ni                            (note: C, P, S not considered here)

Melilite is a binary solid solution, exhibiting complete solid solution between the end-members gehlenite and åkermanite. It is a charge-coupled substitution, since 2 $Al^{3+}$ are replaced by the $Mg^{2+}$ - $Si^{4+}$ pair. Mg-olivine has limited substitution of Ca, but complete substitution of Fe. Efforts to account for such crystal-chemical effects in equations of state comprise a very large geological, metallurgical, and materials science literature, both experimental and theoretical (*Geiger*, 2001). For example, existing models are not capable of accurately predicting the crystallization of melilite from silicate liquid (*L. Grossman et al.*, 2002, their Fig. 1), therefore they are unlikely to model gas-melilite equilibria correctly either. Yet there is no guarantee that tweaking the model for melilite thermodynamics to work with a particular liquid model will improve the situation. Consistency of calibration across all the models used in a calculation is a goal of future work.

   The most successful efforts at modeling groups of pure end-member minerals (e.g., *Berman*, 1988), mineral solid solutions (e.g., *Sack and Ghiorso*, 1989), or complex geochemical systems such as crystallizing magmas (e.g., *Berman*, 1983; *Ghiorso and Sack*, 1995), are based on simultaneous optimization of equations of state that describe multiple experimentally determined equilibrium phase relations. Results of these efforts to describe terrestrial rocks are called 'internally consistent', because each piece of data is evaluated and weighted in the context of the entire data set to optimize agreement with measurements. Less data is available to describe less common substitutional elements (e.g.-Cr, Mn, Ni in olivines and pyroxenes) which may be more important in extraterrestrial rocks.

   The most complex solid solution important to the condensation of rocky material at high T is Ca-rich pyroxene. A comprehensive formulation of a thermodynamic model of pyroxenes with the general formula:

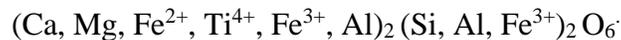

$$(Ca, Mg, Fe^{2+}, Ti^{4+}, Fe^{3+}, Al)_2 (Si, Al, Fe^{3+})_2 O_6.$$

is by *Sack and Ghiorso* (1989; 1994), who concentrated their calibration range on the pyroxene compositions common to basaltic igneous rocks. In particular, their Ti-, Al-pyroxene properties were constrained mainly by studies of basaltic FeO-bearing systems (*Sack and Carmichael*, 1984). Their model reproduces many of the chemical characteristics of experimental and natural pyroxene-bearing mineral assemblages, but not the $Ti^{3+}$-bearing pyroxenes (formerly known as fassaite) found in CAIs. *Beckett* (1986) identified four endmembers critical to describing pyroxenes in CAIs: diopside = $CaMgSi_2O_6$, CaTs = $CaAl_2SiO_6$-, $T_3P$ = $CaTi^{3+}AlSiO_6$, and $T_4P$ = $CaTi^{4+}Al_2O_6$; estimated Gibbs free energies for the Ti species; and provided a robust method for calculating these component molecular fractions from oxide weight percent data. Although it includes the $Ti^{3+}$ substitution, the model used by *Yoneda and Grossman* (1995) ignores Fe, and is not calibrated against models of other solid solutions, or against liquids.

A thermodynamic activity model for solid metal alloy was developed by *Grossman et al.* (1979) and revised by *Ebel and Grossman* (2000). Although treatments of more complex systems are reported in the metallurgical literature (e.g., *Fernández-Guillermet*, 1988), they have not yet been incorporated into published condensation calculations. Implementations used in industry are proprietary (e.g., *Eriksson*, 1975; *Eriksson and Hack*, 1990; *Bale et al.*, 2002; http://www.esm-software.com/chemsage/). Other workers have treated specific metal subsystems (e.g., *Wai and Wasson*, 1979; *Fegley and Palme*, 1985). *Kelly and Larimer* (1977) used thermodynamic theory to trace the history of siderophiles from condensation through planetary differentiation, however, the effects of C, S, and P were not considered. *Petaev et al.* (2003) have coupled a model for diffusion in metal (*Wood*, 1967) with condensation chemistry to address the formation of zoned metal grains (*Meibom et al.*, 2001). The results described here do not include any assumptions about grain sizes or residence times, nor is diffusion data considered strong enough to warrant modeling diffusion in large-scale dust-gas systems.

In reduced chemical systems with high C/O ratios, carbide, nitride, and sulfide minerals are calculated to be stable, and also intermetallic compounds (e.g., FeSi). We do not yet know how to model many reduced solid solutions. For example, a wide range of niningerite (Fe, Mg, Mn)S compositions is found in meteorites (*Ehlers and El Goresy*, 1988), yet this solid solution is not accounted for in existing equilibrium stability calculations for highly reducing systems (e.g., *Lodders and Fegley*, 1997). This is a fertile area for future work.

**2.2.6. Liquid Solutions.** Two well-tested models exist for describing the thermodynamic activities of oxide components in silicate liquids, suitable for modeling crystal-liquid equilibria in magmatic systems. Igneous CAIs and chondrules are subsets of the universe of such systems. Both models use a classical non-ideal thermodynamic formalism (cf., *Anderson and Crerar*, 1993) to describe the chemical potential energy of silicate liquids and are calibrated against internally consistent data sets describing coexisting mineral phases (but not metal alloy). *Berman's* (1983) model is restricted to $CaO-MgO-Al_2O_3-SiO_2$ (CMAS) liquids. *Ghiorso and Sack's* (1995) 'MELTS' model includes additional components containing the oxides $TiO_2$, FeO, $Cr_2O_3$, $P_2O_5$, $Na_2O$, and $K_2O$, but because of their choice of stoichiometric end-member components, it requires that liquids have molar $SiO_2 > [\frac{1}{2} (MgO-Cr_2O_3) + \frac{1}{2} FeO + (CaO-3 P_2O_5) + Na_2O + \frac{1}{2} K_2O]$. In the absence of a tested, universal liquid model, one must use either one model or the other, depending upon the composition region of interest (cf., *Ebel and Grossman*, 2000; *Ebel*, 2005). *Berman's* (1983) model can be applied to CAIs and FeO-poor chondrules, while the MELTS model applies to most chondrule compositions. The limit on MELTS liquid compositions is particularly restrictive in considering FeO evaporation (*Ebel*, 2005).

To calibrate the thermodynamic parameters of a model liquid requires multiple statements of crystal-liquid equilibria at known P and T derived from laboratory data, combined with accurate thermodynamic models of the crystalline phases involved (*Ghiorso*, 1985, 1994; *Berman and Brown*, 1987). The *Berman* (1983) CMAS model is calibrated using an optimized end-member mineral thermodynamic data set (*Berman*, 1983), but all solid solutions are treated as mechanical mixtures of end-members (e.g., *L. Grossman et al.*, 2002, their Fig. 1). The MELTS model (*Ghiorso and Sack*, 1995) is calibrated using an internally consistent end-member data set for solid mineral end-members (*Berman*, 1988), and complex, non-ideal solid solution models for olivine (*Sack and Ghiorso*, 1989), pyroxene (*Sack and Ghiorso*, 1994), feldspar (*Elkins and Grove*, 1990), spinel (*Sack and Ghiorso*, 1991a, b), etc. These models include energetic effects of atomic order/disorder in crystalline lattices, but not magnetic effects that may contribute below ~800 K. In calculations involving solid solutions and liquids, there are many choices to be made; none is perfect.

**Table 1:** Names, abbreviations and chemical formulae of mineral phases.

| mineral | abbrv. | chemical formula |
| --- | --- | --- |
| corundum | cor | $Al_2O_3$ |
| hibonite | hib | $CaAl_{12}O_{19}$ |
| grossite | grs | $CaAl_4O_7$ |
| Ca-monoaluminate | CA1 | $CaAl_2O_4$ |
| perovskite | prv | $CaTiO_3$ |
| melilite | mel | $Ca_2(Al_2, MgSi)SiO_7$ |
| Al-spinel | Al-spn | Al-rich $(Fe,Mg,Cr,Al,Ti)_3O_4$ |
| Cr-spinel | Cr-spn | Cr-rich $(Fe,Mg,Cr,Al,Ti)_3O_4$ |
| olivine | olv | $(Mg_2, Fe_2, MgCa)SiO_4$ |
| metal alloy | met | Fe,Ni,Co,Cr,Si |
| feldspar | fsp | $(CaAl, NaSi, KSi)AlSi_2O_8$ |
| Ca-pyroxene | Ca-px | $Ca(Mg,Fe,Ti^{4+},Al,Si)_3O_6$ |
| orthopyroxene | opx | $MgSiO_3$-$FeSiO_3$ |
| rhombohedral oxide | rhm oxide | $(Mg,Fe^{2+},Ti^{4+},Mn)_2O_3$ |
| pyrophanite | - | $MnTiO_3$ |
| whitlockite | wt | $Ca_3(PO_4)_2$ |
| troilite | - | FeS |
| oldhamite | - | CaS |
| osbornite | - | TiN |
| graphite | - | C |
| cohenite | - | $Fe_3C$ |
| *Suppressed:* | | |
| cordierite | - | $Mg_2Al_4Si_5O_{18}$ |
| sapphirine | - | $Mg_4Al_8Si_2O_{20}$ |

Formulae are restricted to solid solution ranges present in calculations (cf., *Ebel and Grossman*, 2000). Ferric iron ($Fe^{3+}$), although present in some solution models (spinel, Ca-px, rhm-ox), is insignificant in all calculated results shown here.

## 2.3. Algorithms

**2.2.1. Gas-Solid-Liquid Equilibria.** In all the calculations considered here, the gas is considered as the primary reservoir of material. Two approaches may be taken. First, the entire system can be solved as a whole, which amounts to treating each pure solid species as if it were a gaseous molecule, but with a very different equation of state. *Yoneda and Grossman* (1995) were the first to incorporate true solid solution behavior into such an approach. Alternatively, *Ebel and Grossman* (2000) solved speciation in the gas separately (cf., *White et al.*, 1958; *Grossman and Larimer*, 1974; *Lattimer et al.*, 1978; *Smith and Missen*, 1982; *Lewis*, 1997), treating the gas as a separate solution phase similar to pyroxene or liquid. These approaches yield identical results for the same inputs. *Ebel and Grossman* (2000) adapted algorithms of *Ghiorso* (1985, 1994) to facilitate use of the solid and liquid solution models embodied in the MELTS code (*Ghiorso and Sack*, 1995). They substituted gas for liquid as the material reservoir in the MELTS algorithm, and treated liquids analogously to solid solution phases.

   ***2.2.2. Solid Stability and Optimization.*** This is the most laborious part of condensation calculations, particularly when a rigorous treatment of solid solutions is included. The goal is to find the optimal distribution of elements between gas and all possible condensates, with no *a priori* knowledge of what the stable condensates are. At any step in a particular calculation, the thermodynamic stability of all potential condensates can be calculated from their equations of state, given the partial pressures of their constituent elements in the gas. Each stable condensate must be added to the system in a small mass increment, with complementary subtraction of mass from the gas. Minimization of the chemical potential energy (e.g., Gibbs energy) of the gas + condensate system follows, by redistribution of matter between all the coexisting phases. Thermodynamic stability of potential condensates is then tested again. Condensates whose mass decreases below a threshold are returned to the gas reservoir (cf., *Ebel et al.*, 2000), where their component elements are removed into more stable condensates in subsequent steps.

## 3. RESULTS

   Results are presented here in terms of total pressure ($P^{tot}$, bar) and temperature (T, in Kelvin) (e.g., Plate 1). Abbreviations and chemical formulae used in Plate 1 and elsewhere are listed in Table 1. Discussion logically follows the behavior of cooling systems of particular bulk composition at fixed $P^{tot}$, following descending vertical paths. It must be emphasized that temperature is only *one* state variable affecting stable phases; pressure is just as important, as is the bulk vapor composition. Boundaries mark the conditions at which particular condensed phases either appear (become stable), or disappear along such cooling paths. Stable mineral assemblages, in *phase fields*, are listed from most to least refractory phase where space permits (e.g., Plate 1). In all cases, vapor is present, and it should be remembered that at most temperatures shown, major elements are continually condensing from gas into solid and/or liquid phases with cooling: the condensate assemblage, alone, does *not* represent a closed system at any point in any diagram.

   Although models for solid solutions (Table 1) contain many potential compositions, less refractory endmembers are not commonly stable in the systems investigated here. Feldspar is nearly always pure anorthite $CaAl_2Si_2O_8$, olivine pure forsterite $Mg_2SiO_4$ with minor Ca. Melilite and Ca-pyroxene show the greatest variation in composition at high temperatures (cf., *Ebel and Grossman*, 2000).

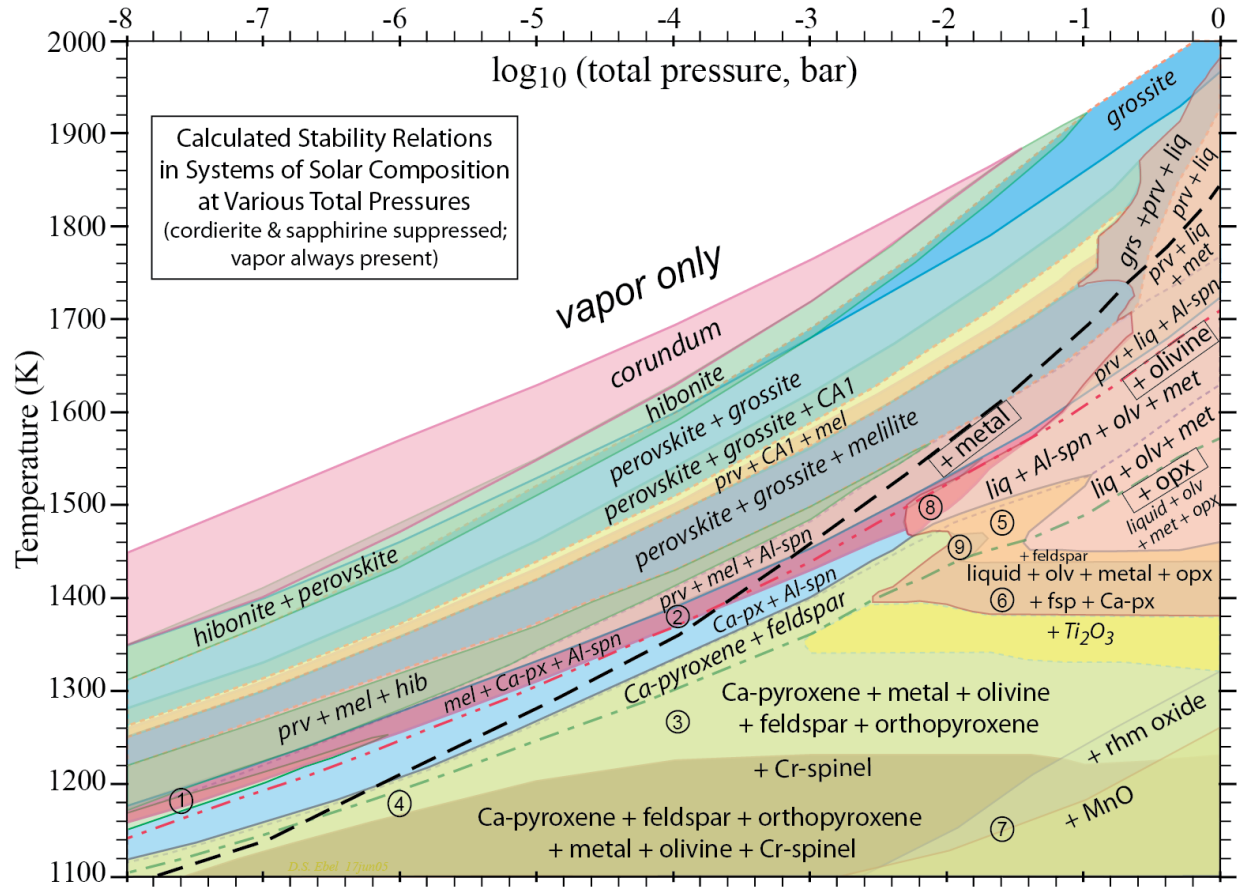

**Plate 1:** Stability of condensed phases with temperature and pressure, in a vapor of solar composition. The temperatures below which metal, olivine, and orthopyroxene are stable are shown as a function of $P^{tot}$ by dashed, dash-dot-dot, and dash-dot lines, respectively. At $P^{tot} > \sim 10^{-2.5}$ bar, liquids, primarily of CMAS composition, become stable. The feldspar here is nearly pure anorthite (alkali-free). Even at high $P^{tot}$, silicates contain almost no FeO-bearing component, in contrast to Mg-silicates at high dust enrichments in Plate 4. Circled numerals illustrate features of this result: At ①, Ca-feldspar is stable at $P^{tot} < \sim 10^{-7.6}$ bar over a small T-range above its persistent T, which is about 80° lower (see Plate 2). At ②, melilite, Ca-pyroxene, and Al-spinel are stable, and melilite will disappear, and olivine will become stable, ~15° above the appearance of metal, as the system cools. The assemblage at ② is commonly observed in melted CAIs. Upon cooling of the system at ②, large quantities of olivine and metal will condense, with Ca-pyroxene and Al-spinel. At the same $P^{tot}$, feldspar will replace Al-spinel at lower temperatures, then orthopyroxene becomes stable, yielding the assemblage at ③, which is the same assemblage as at ④. Phases stable at ⑤ are liquid + olivine + metal + feldspar, with orthopyroxene and then Ca-pyroxene condensing at successively lower T to yield assemblage ⑥, in which all liquid crystallizes to solids at ~1380 K. Olivine + melilite + Al-spinel + liquid + metal are stable at ⑧, and metal + olivine + feldspar + Ca-pyroxene + liquid at ⑨. Assemblages ③-⑥ are relevant to FeO-poor chondrules (e.g., *Jones and Scott*, 1989). The oxide $Ti_2O_3$ is predicted to form when liquid crystallizes, however this Ti should, in real systems, dissolve into silicates or oxides, as should the MnO predicted to condense at ⑦ at $P^{tot} \sim 10^{-1.6}$. These 'shoulds' stem from the extreme difficulty of accurately calibrating solid solution models of minor components.

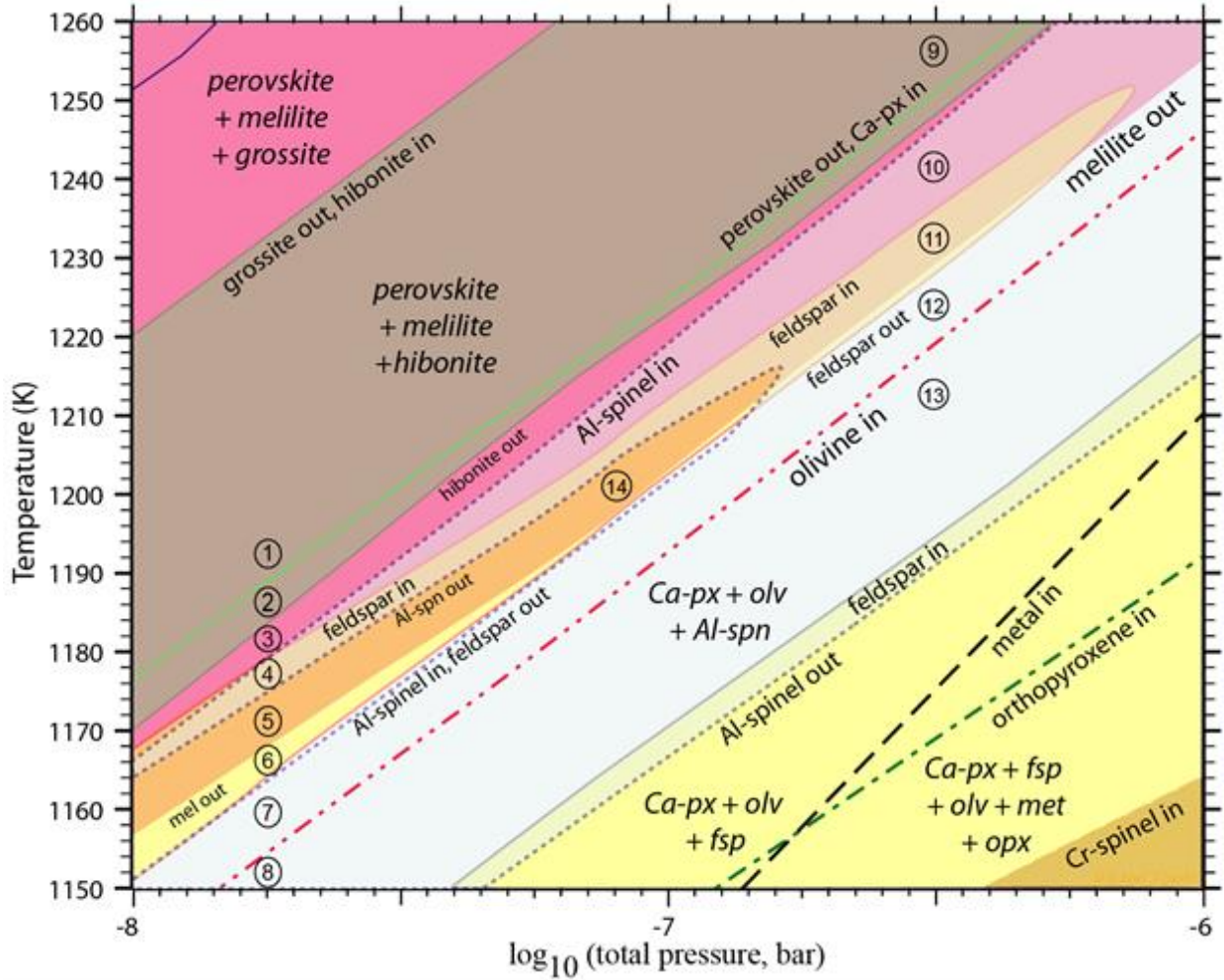

**Plate 2:** Stability of feldspar at low total pressure, in a vapor of solar composition. This is a detail of Plate 1. The stable solids at $P^{tot} = 10^{-7.75}$, as Si and Mg condense from vapor as the system cools are: ① perovskite + melilite + hibonite; ② melilite + hibonite + Ca- pyroxene; ③ melilite + Ca- pyroxene; ④ melilite + Ca- pyroxene + Al-spinel + feldspar; ⑤ melilite + Ca- pyroxene + feldspar (as at (14)); ⑥ Ca- pyroxene + feldspar; ⑦ Ca- pyroxene + Al-spinel; ⑧ Ca-pyroxene + Al-spinel + olivine. At $P^{tot} = 10^{-6.5}$, as the system cools, assemblages are: ⑨ perovskite + melilite + hibonite; (10) melilite + Ca- pyroxene + Al-spinel; (11) melilite + Ca-pyroxene + Al-spinel + feldspar; (12) Ca- pyroxene + Al-spinel + feldspar; (13) Ca- pyroxene + Al-spinel + feldspar + olivine.

## 3.1. Equilibria in a Gas of Solar Composition

### 3.1.1. Effect of Pressure.
Plate 1 illustrates equilibrium stability relations of vapor, solid, and liquid phases calculated by this author, with $10^{-8} < P^{tot} < 1$ bar, and $1100 < T < 2000$ K, in a closed system of solar composition (*Anders and Grevesse*, 1989). The method and data are those reported in *Ebel and Grossman* (2000), who found that the refractory minerals cordierite and sapphirine should become stable by reaction with other solids at T and $P^{tot}$ below the temperature at which virtually all Mg, Al, Si, and Ca are condensed. Cordierite and sapphirine are extremely rare in meteorites (*Sheng et al.*, 1991; *Rubin*, 1997), they are calculated to be stable only at relatively low temperatures, and assemblages bearing them are only stable, relative to those reported here, by order $< 10^{-2}$ J (e.g., at $P^{tot}=10^{-4}$, $8x10^{-4}$ J at 1300 K, $3x10^{-3}$ at 1200 K) relative to other solids (i.e., well within error); they have been omitted from all the calculations reported here. Although Na-, Fe-free cordierite is calculated to be near the minimum free energy configuration, the cordierite found by *Fuchs* (1969) has 3-6 wt% $Na_2O$. The absence of cordierite in meteorites suggests the importance of mineral reconstructive kinetics even at temperatures near the stability field of feldspar.

Plate 2 shows a sub-portion of Plate 1, illustrating phase relations involving feldspar at low $P^{tot}$. Phase boundaries are marked 'in'/'out' to signify phase appearance/disappearance upon cooling at constant $P^{tot}$. A common occurrence in this and other results (e.g., liquid field in Plate 1) is the stability of a particular phase over two different T ranges at a fixed $P^{tot}$. This results from the continued condensation of major elements (e.g., Si) upon cooling (cf., *Yoneda and Grossman*, 1995, liquid field in their Fig. 5). The condensate assemblages continually adjust so that the chemical potential energy of the system is minimized. Feldspar is nearly pure anorthite. Textures in CAIs can be interpreted to show that anorthite was thermodynamically stable at higher temperatures than olivine and other phases, inconsistent with calculated phase diagrams. *MacPherson et al*. (2004) discussed this phenomenon in the context of calculations showing anorthite condenses at a higher T than forsterite below $P^{tot} \sim 10^{-4}$ bar (their Fig. 2), with dramatic effect (their Fig. 3). Ca-pyroxene is stable in all these assemblages. For the calculation by *MacPherson et al*. (2004) to predict the correct phase relations, the endmember pyroxene energies of *Sack and Ghiorso* (1994) must be innaccurate, but must be fortuitously corrected by *MacPherson et al*. (2004) ignoring the mixing properties included in the pyroxene model of *Sack and Ghiorso* (1994; see Section 3.1.2). Results of a calculation using the MELTS implementation of that model (Plate 2) illustrate that a narrow high-T anorthite stability field does exist at very low $P^{tot} < \sim 10^{-6.8}$. If feldspar (anorthite) were to form in this T-P range, then it might survive metastably, relative to other phases, in a cooling system. Other arguments for anorthite stability have been discussed by *Weisberg et al.* (2005).

### 3.1.2. Comparison with Other Results.

Several recent calculations of equilibrium condensation are compared in Fig. 1. Temperatures of mineral phase appearance and disappearance are shown for three specific $P^{tot}$ (bar) published as tables by *Wood and Hashimoto* (1993) and *Petaev and Wood* (1998b), who did not include Ti or noble gases; *Yoneda and Grossman* (1995), who included Ti and He but not Ne or Ar; *Petaev and Wood* (2000), with Ti but no noble gases; *Lodders* (2003) who included the entire periodic table, but did not publish disappearance temperatures of phases; and *Ebel and Grossman* (2000), whose methods were the same as this work. Elemental abundances are from *Anders and Grevesse* (1989) in all cases except for *Lodders* (2003). Inclusion or absence of noble (inert) gases slightly changes the effective $P^{tot}$ in the calculation. The absence of Ti strongly affects phase stability, particularly by preventing perovskite formation. There are many subtle comparisons that could be made using these results, however it is their overall similarity, particularly for the major phases found in chondrites, that is striking.

All three groups used different data for the chemical potential energy (Gibbs free energy) of endmember mineral phases, with varying degrees of documentation. Differences between *Yoneda and Grossman* (1995) and earlier work are detailed in that paper, and differences between that work and *Ebel and Grossman* (2000), and by extension the present work, are explained in detail in the (2000) paper. In particular, the subtle equilibria among the Ca-Aluminates (CA1, hib, grs) among the papers by Grossman's group have all been documented to result from choices of basic data made for internal consistency with liquid solution models (*Ebel and Grossman,* 2000). The other major difference between *Ebel and Grossman* (2000), this work, and that of *Petaev and Wood* (2000), appears to result from the formers' use of the complete MELTS (*Sack and Ghiorso*, 1994) calculation engine for Ca-pyroxene. *Petaev and Wood* (1998a,b; 2000) use endmember data from, e.g., *Sack and Ghiorso* (1989, 1994), but they used ideal mineral solution models for the mixing properties of mineral solid solutions between those endmembers (*Petaev and Wood*, 2000). *Lodders* (2003) did not consider solid solutions, so, for example, her 'melilite' is pure gehlenite. For purposes of parameterizing condensation for astrophysical calculations, all of these results are adequate, and that of *Lodders* (2003) is preferred for its complete coverage of all the (astrophysical) 'metals' (see Section 5.1). The work of *Ebel and Grossman* (2000), and the present calculations, are most consistent with models of liquid solutions at conditions where the latter are thermodynamically stable, and include full implementation of the internally consistent MELTS models for crystalline solutions that coexist with mafic (Mg-Fe-rich multicomponent silicate) liquids.

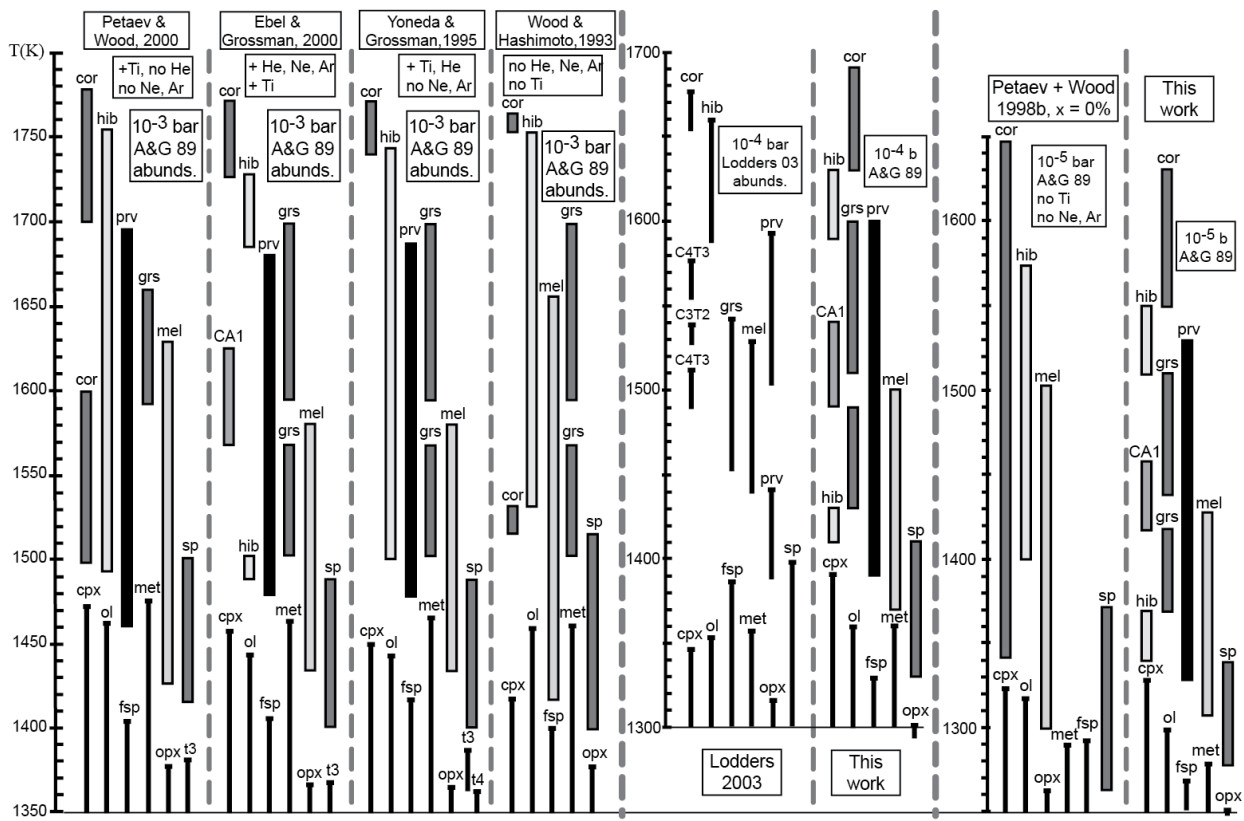

**Figure 1:** Comparison of published results for vapor of solar composition (see text 3.1.2). Mineral abbreviations are: t3=Ti$_3$O$_5$, t4=Ti$_4$O$_7$, C4T3=Ca$_4$Ti$_3$O$_{10}$, C3T2=Ca$_3$Ti$_2$O$_7$, sp=Al-spinel, cpx=Ca-pyroxene, and as listed in Table 1.

## 3.2. Equilibria in Carbon-enriched Systems

Reduced phases are common constituents of enstatite chondrites (e.g., CaS, MgS, etc.; *Keil*, 1968, 1989; *Brearley and Jones*, 1998). Graphite, SiC, and TiC are common pre-solar grains (*Meyer and Zinner*, 2005). Both hydrous and anhydrous interplanetary dust particles (IDPs) are suffused by carbon-rich intergranular material (*Flynn et al.*, 2003), and CI chondrites contain up to 5 wt% carbon (*Lodders and Osborne*, 1999). A diversity of carbonaceous material was abundantly present in the protoplanetary disk (*Kerridge*, 1999), so exploration of mineral equilibria in hot, C-enriched systems of otherwise solar composition has been an active area of research (*Larimer*, 1975; *Larimer and Bartholomay*, 1979; *Wood and Hashimoto*, 1993; *Sharp and Wasserburg*, 1995; *Lodders and Fegley*, 1993, 1998).

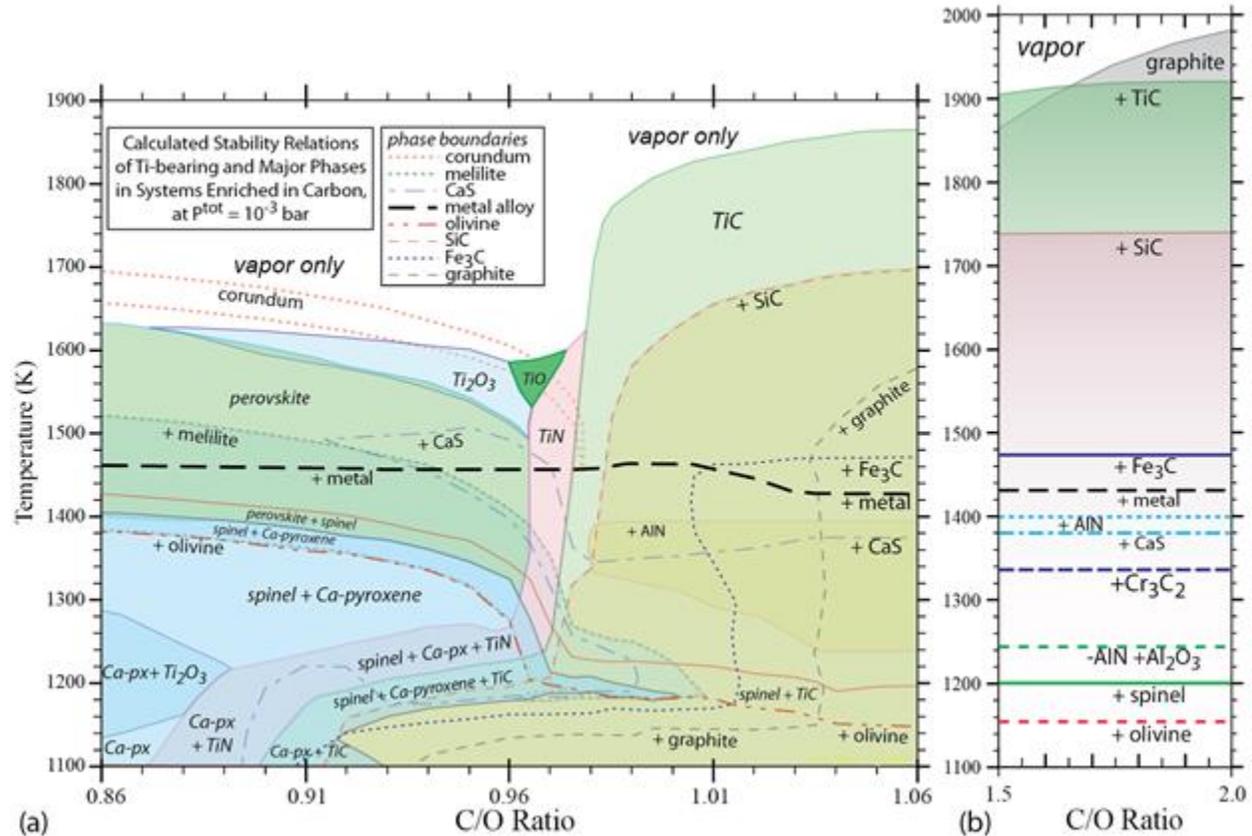

**Plate 3: a)** Stability of Ti-bearing phases, and selected other phases, in a vapor of solar composition with C added to increase the C/O ratio. Detail of phase stability at high C/O ratio, as excess C condenses as graphite. Circled numbers in Plate 3a are: At ①, C/O~0.967, 1400 K, CaS (oldhamite) is stable, with TiN (osbornite), as well as hibonite and metal alloy. At ②, 120° lower, Ti is in spinel solid solution, TiN, and perovskite, and Ca is in melilite. Melilite is stable over a successively smaller T region up to C/O ~1.01. At ③, C/O=0.905, 1180 K, Ti is in spinel, Ca-pyroxene and TiN, coexisting with olivine, metal, and CaS. At ④ and ⑥ Ti is in spinel and TiC, while at ⑤ all the Ti is in TiC. Metal coexists with $Fe_3C$ (cohenite) at all three points. At ④, olivine and CaS are also stable; at ⑤, SiC and AlN; and at ⑥, olivine, spinel, SiC, CaS, graphite, and $Cr_3C_2$. Field boundaries for AlN and $Cr_3C_2$ are not shown.

Plate 3a illustrates stability relations among Ti-bearing phases, with other selected phases, shown in a narrow range of C/O where equilibrium chemistry changes dramatically (cf., *Lodders and Fegley*, 1997, their Fig. 3). At C/O < 0.85, equilibria are grossly similar to C/O ~ 0.43 (solar). Only Ti-bearing phases are listed in most fields, while phase boundaries of Ti-free phases *x* upon cooling at constant C/O are labeled "+*x*". The calculations were done using the data and methods of *Ebel and Grossman* (2000). As the C/O ratio approaches unity, reduced phases become stable (e.g., CaS, TiN, TiC). At C/O > 1, carbides and graphite are stable at the highest temperatures, and graphite condenses at progressively higher T as C/O increases (Plate 3b). Oxides (e.g., Mg-olivine, spinel) are also stable at high C/O, because condensation of graphite liberates O from gaseous CO. Metal becomes much more Ni-, Si-, and Co-rich than under oxidizing conditions. Intermetallic Fe-Si alloys are not treated rigorously in the calculation, but information on these compounds (e.g., FeSi, FeSi$_2$) does exist (e.g., *Meco and Napolitano*, 2005). Unfortunately, while some data exist on binary solutions, data is sparse regarding the thermochemical behavior of more complex reduced solid solutions (e.g., (Ti, Si, Zr)C, (Ti, Al, W)N, (Ca,Fe,Mg,Mn)S).

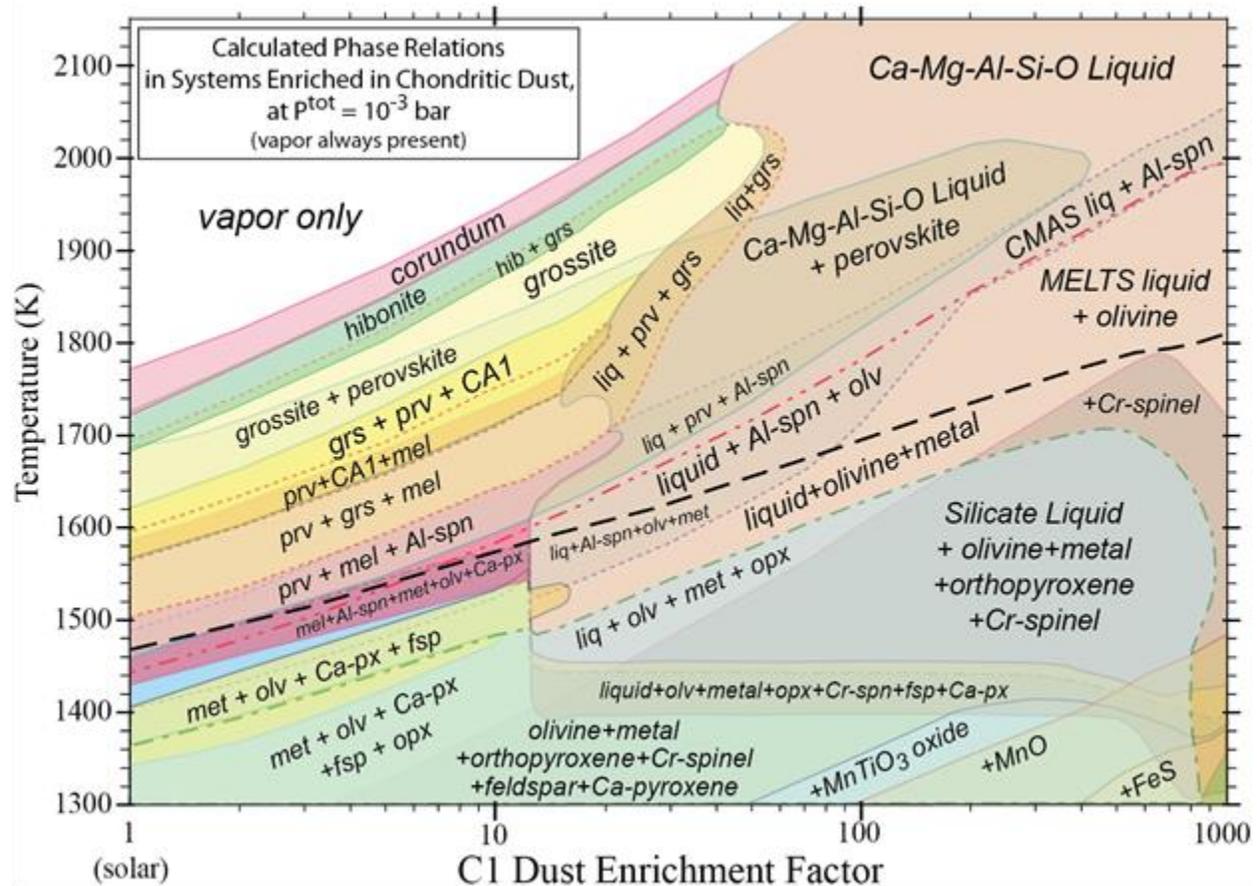

**Plate 4:** Calculated phase relations in systems enriched in a CI-chondrite-like dust at a total pressure (P$^{tot}$) of 10$^{-3}$ bar, calculated as in *Ebel and Grossman* (2000). See Table 1 for formulae and abbreviations. Circled numbers are assemblages: ① perovskite + Ca-monoaluminate + liquid, ② perovskite + Ca-monoaluminate, ③ melilite + Al-spinel + metal + Ca-pyroxene, ④ Al-spinel + metal + olivine + Ca-pyroxene, ⑤ Al-spinel + metal + olivine + Ca-pyroxene + feldspar, ⑥ liquid + olivine + metal + feldspar, ⑦ liquid + olivine + metal + orthopyroxene + feldspar + Ca-pyroxene, ⑧ liquid + olivine + metal + orthopyroxene + Cr-spinel + feldspar, ⑨ liquid + olivine + metal + Cr-spinel + MnO, ⑩ liquid + olivine + metal + Cr-spinel + feldspar + MnO, (11) =⑩ + Ca-pyroxene, (12) =(11) + sulfide + MnTiO$_3$, (13) = (12) + whitlockite, (14) Ti$_3$O$_5$ is briefly stable in this field. Manganese is calculated to condense as MnO, as well as MnTiO$_3$ oxide, that would both dissolve in real coexisting silicates (e.g., olivine, pyroxene).

### 3.3.  Equilibria in Dust-enriched Systems

Some mixture of dust grains is expected to accumulate at the midplane of the protoplanetary disk, and this dust may be the precursor material for chondrule and/or CAI formation. *Ebel and Grossman* (2000) investigated enrichment of the solar composition in a dust approximating the composition of the meteorite Orgueil (*Anders and Grevesse*, 1989). Chondritic dust is a logical assumption, given the chondritic abundances observed in bulk protoplanetary material. Various other kinds of dust can be supposed, for example a C-rich dust similar in composition to unequilibrated, anhydrous, interstellar organic- and presolar-silicate-bearing cluster interplanetary dust particles (*Ebel and Alexander*, 2005). Local vaporization of micron-sized dust during rapid heating events (e.g., *Desch and Connolly*, 2002; *Joung et al.* 2004), might affect chondrule composition by suppressing evaporation (*Galy et al.* 2000; *Alexander et al.*, 2000).  Vaporized dust would promote pre-accretionary metasomatic alteration, particularly affecting chondrule mesostasis (*J. Grossman et al.*, 2002), or forming refractory alkali-bearing minerals in CAIs (e.g., sodalite; *Allen et al.*, 1978; *McGuire and Hashimoto*, 1989; *Krot et al.*, 1995; *Nagahara*, 1997).

Plate 4 illustrates the assemblages of solid and liquid phases that are calculated to be thermodynamically stable relative to vapor, at fixed total pressure, $P^{tot} = 10^{-3}$ bar, as functions of temperature (K) and enrichment in chondritic dust (cf., *Ebel and Grossman*, 2000). The oxides become thermodynamically stable at progressively higher temperatures with increasing dust enrichment, leading to the stability of liquids, and substantial FeO dissolved in silicates (*Ebel and Grossman*, 2000, their Fig. 8). Detailed results such as calculated oxygen fugacity, FeO content of silicates, composition of metal alloy, and melilite composition are presented in *Ebel and Grossman* (2000) for dust enrichment increments of 100x. Cordierite and sapphirine are calculated to be stable below 1330 K, at dust enrichments < 20x, but they are entirely omitted from this calculation for reasons detailed above. Calculations to even higher dust enrichments, essentially for a pure chondritic dust, were presented by *Ebel* (2001).

## 4.  METEORITIC EVIDENCE FOR CONDENSATION

### 4.1.  Volatility-related Differences Between Chondrites

Bulk compositions of chondrites are reviewed in detail by *Hutchison* (2004), who offers a somewhat strained dichotomy of causal models for differences between them (his Table 7.1). He concludes that "the chemical compositions of the classes and groups of chondrites were established before chondrule formation. Elemental fractionation correlates with volatility..." , when CI chondrites are considered to be the most primitive, basic composition, due to their closest equivalence to the composition observed in the solar photosphere (*Hutchison*, 2004, p.188). Regarding the constituents of chondrites, *Chiang* (2002) characterizes chondrule (and CAI) formation as "a zeroth order" unsolved problem. *Hutchison* (2004, p.192) considers the problem "even more intractable and emotive than identifying the mechanisms for fractionating CI chondrite material to make the chondrite groups". Chondrite compositions are not simply additive combinations of Mg-Si chondrules, metal, and Ca-, Al-rich inclusions. Instead, some complementarity must be invoked, in which chondritic material was fractionated, based on relative volatilities of elements, and then chondrules, etc., formed from these fractionated batches (*Hutchison*, 2004, p. 193; *Palme*, 2001).

One approach to deciphering the elemental differences among the chondrites, and comparing them to planetary bodies, is to assume that volatile elements were as discrete components. An obvious first choice is water ice ($H_2O$). Figure 2a illustrates the abundances of selected elements (atomic fraction of total; calculated from data of *Wasson and Kallemeyn* (1988; cf., *Hutchison*, 2004), with all H and O = ½ H removed. The relative abundances of C and S are still very high in CI and CM chondrites, relative to the other classes (cf., *Weisberg et al.*, 2006), with complementary depletion in Si, Mg, and Fe. Relatively volatile alkalies, and Se and Ga, are enriched in the same classes. Large variations in Fe are seen in the enstatite chondrites, particularly EH, and these are reflected in siderophile Co, Ga, and P, and chalcophile Se. These data suggest that iron alloy fractionation, perhaps even prior to chondrule formation, contributed to the differences between these classes.

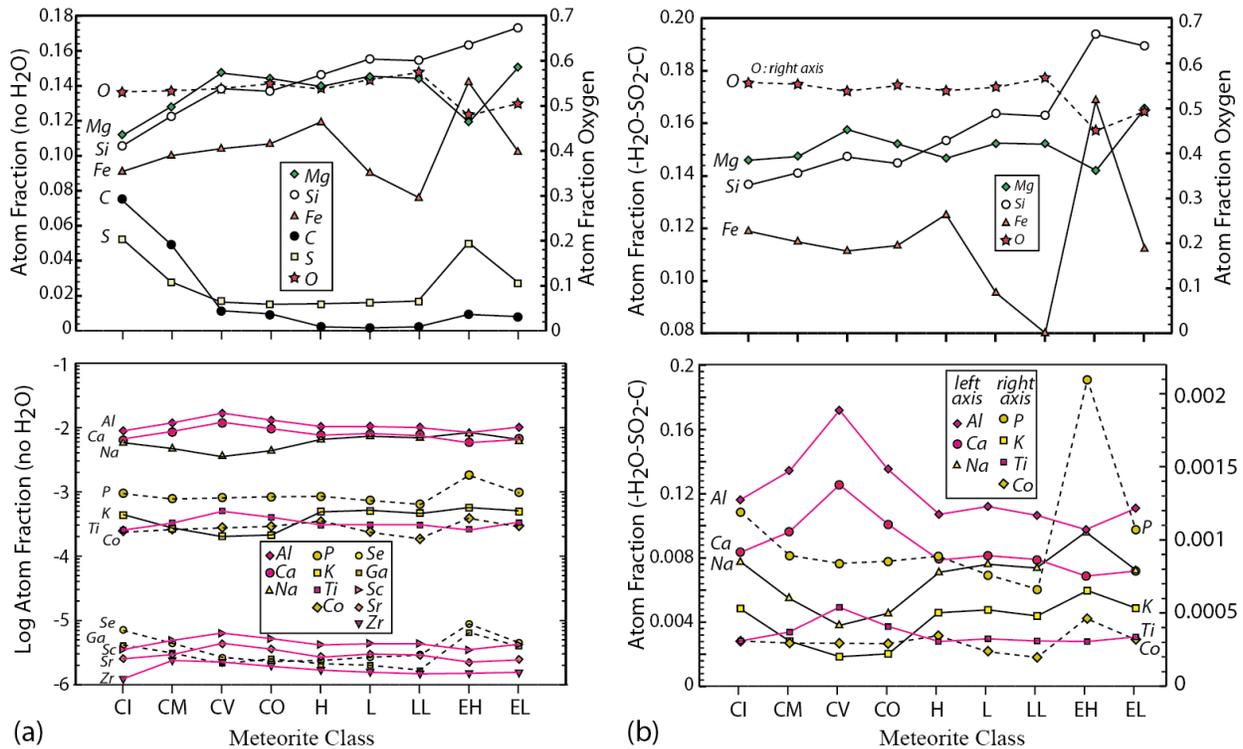

**Figure 2:** Volatility-related elemental fractionation among chondrite groups, for selected elements with different cosmochemical affinities. (a) All H and O=0.5*H subtracted. (b) All C and S (as $SO_2$) subtracted.

Figure 2b shows selected element abundances, recalculated after subtraction of all C as pure carbon (consistent with studies of interplanetary dust), and all S as $SO_2$, which may be less valid because it is not clear how important FeS was as a chondrite precursor. This recalculation removes most bias in illustrating the enrichment of CV and CO chondrites in highly refractory Al, Ca, Ti, Sc and Zr. This is the major volatility-related fractionation seen in carbonaceous chondrites and extends to other highly refractory elements not shown here. The variation in alkalies (Na, K) is also likely to be related to their volatility. Variation in the siderophile elements (Fe, Co, P) is most pronounced among the ordinary (H, L, LL) and enstatite (EH, EL) chondrites. The cause of abundance variations of the siderophile elements among chondrites is not clear to this author. Loss (or gain) of metal grains, carrying siderophile elements in chondritic relative abundances, has been hypothesized, but no cause is clearly established. More volatile chalcophile elements ("S-loving", e.g., Se, Sb) would go along with these metal grains, if they were condensed with them. Alternatively, some fraction of the chalcophile component, and also of similarly volatile elements up to the temperatures of stability of solids containing alkalies, was added to the chondritic components (chondrules, CAIs, metal grains, matrix) before (*Allen et al.*, 1978) or during the components' accretion into meteorite parent bodies.

## 4.2. Meteorite Components

A wide variety of inclusions and matrix grains is found in meteorites, comprehensively reviewed by *Brearley and Jones* (1998). Many have been hypothesized to be condensates, or to have equilibrated with vapor at high T in the protoplanetary disk or nebula. These include, but are not limited to: (1) 'fluffy' type A (FTA) inclusions (*MacPherson et al.*, 1984; *Beckett*, 1986; but cf., *MacPherson*, 2003), spinel-pyroxene aggregates (*Kornacki and Fegley*, 1984; *MacPherson et al.* 1983), hibonite-bearing objects (*MacPherson et al.*, 1983; *Hinton et al.*, 1988; *Ireland et al.*, 1991; *Russell et al.*, 1998; *Simon et al.*, 1998, 2002; *MacPherson*, 2003), and other highly refractory inclusions (*Bischoff et al.*, 1985); (2) CAIs and chondrules (*Wood*, 1963; *Grossman and Clark*, 1973; *Wänke et al.*, 1974; *Beckett and Grossman*, 1988; *Boss*, 1996; *Lin and Kimura*, 2000; *Scott and Krot*, 2001; *Kurat et al.*, 2002; *Lin et al.*, 2003), particularly those with unfractionated REE abundance patterns (*Russell et al.*, 1998); (3) minerals deposited in voids in type B CAIs (*Allen et al.*, 1978);  (4) amoeboid olivine aggregates (*Grossman and Steele*, 1975; *Komatsu et al.*, 2001; *Aléon et al.*, 2002; *Krot et al.*, 2004a; *Weisberg et al.*, 2005); (5) metal grains in CB and CH chondrites (*Meibom et al.*, 2001; *Petaev et al.*, 2001, 2003; *Campbell et al.*, 2001), and metal in CR chondrules (*Connolly et al.*, 2001) (7)  PGE nuggets (*El Goresy et al.*, 1979; *Palme and Wlotzka*, 1976) or opaque assemblages (a.k.a.-Fremdlinge; *Sylvester et al.*, 1990) in CAIs; (8) relict hibonite (*El Goresy et al.*, 1984) and fassaite (*Kuehner et al.*, 1989; *Lin and Kimura*, 2000) in CAIs and relict olivine (with included pockets of silicate melt) in chondrules and matrix (*Olsen and Grossman*, 1978; *Steele*, 1986, 1995; *Weinbruch et al.*, 2000; *Pack et al.*, 2004); (9) fine-grained components in matrix and inclusion rims (*Wood*, 1963; *Wark and Lovering*, 1977; *Huss et al.*, 1981; *Kornacki and Wood*, 1984; *Brearley*, 1993; *Greshake*, 1997; *Ruzicka*, 1997; *Bischoff*, 1998; *Wark and Boynton*, 2001; *Zolensky et al.*, 2003); (10) xenoliths in Allende (*Kurat et al.*, 1989); (11) FeO-rich matrix olivine (*Weisberg et al.*, 1997; *Weisberg and Prinz*, 1998). Some fraction of interplanetary dust particles (IDPs), probably sourced in comets, may be composed almost entirely of condensates (*Rietmeijer*, 1998).

There is intense debate about what the extinct radioisotopic systematics (*MacPherson et al.*, 1995; *Russell et al.*, 1998, 2005; *McKeegan et al.*, 2000; *Guan et al.*, 2000), O-isotopic signatures (*McKeegan et al.*, 1998; *Young and Russell*, 1998; *Krot et al.*, 2002; *Young et al.*, 2002; *Clayton*, 2004), and refractory mineralogies (*Grossman*, 1975; *Simon et al.*, 1997, 1998, 2002; *Russell et al.*, 1998) of CAIs imply about their formation histories, as reviewed by *MacPherson* (2003) and in this volume. Fractional condensation, condensation followed by gas-phase metasomatism, and gas-solid equilibrium interrupted by melting events all seem to have affected the most refractory inclusions (e.g., *Stolper and Paque*, 1986). For example, condensation and aggregation of solids to make fluffy type A CAIs, followed by reaction with vapor in the Ca-pyroxene stability field, followed by melting, may produce type B CAIs (e.g., *Beckett*, 1986; perhaps at ② in Plate 1). Evaporation of many type B CAIs is evident in their heavy isotope enrichment (*Clayton et al.*, 1988), and the effect of evaporation on these objects is receiving much attention (e.g., *Nagahara and Ozawa*, 2000; *Galy et al.*, 2000; *L. Grossman et* *al.*, 2002; *Richter et al.*, 2002; *Davis*, 2003; *Alexander*, 2004). The precursor dust aggregates of many melted CAIs and chondrules are highly likely to correspond in bulk composition to some assemblages predicted in Plates 1 or 4. Other inclusion compositions (e.g., radiating pyroxene chondrules) require very extraordinary conditions to predict (*Ebel et al.*, 2003).

Chondrules and igneous CAIs can be thought of as part of a continuum including Al-rich objects (*Ebel and Grossman*, 2000, their Figs 16, 17; *MacPherson*, 2003; *MacPherson and Huss*, 2005). Principal component analyses (*Grossman and Wasson*, 1983) and other evidence (*Bischoff and Keil,* 1984; *Weisberg and Prinz*, 1996) indicate these inclusions formed by rapid melting of mixtures of precursor components. The precursors included refractory, Ca, Al-rich material; less refractory Mg-, Si-rich material; metal alloy and/or metal sulfide; and a more volatile, Na- and K-bearing component (*Grossman*, 1996). Some CAIs may bear trace element signatures that were established during stellar condensation of their precursor grains (*Ireland*, 1990). The sizes and relative proportions of inferred precursors were highly variable, and some may be preserved as relict grains, for example olivine (e.g., *Steele*, 1986; *Jones*, 1996; *Pack et al.*, 2004, 2005), and spinel (e.g., *Misawa and Fujita*, 1994; *Maruyama and Yurimoto*, 2003) in chondrules (cf., *Misawa and Nakamura*, 1996); perovskite in CAIs (e.g., *Stolper*, 1982*; Beckett*, 1986). From a cursory inspection of Plates 1 or 4, these precursor components are consistent with progressive, volatility-controlled fractionation. The very existence of CAIs, chondrules, and more volatile-rich matrix demonstrates that such a fractionation process isolated these objects from thermochemical interaction in some locations prior to their accretion into chondrites.

Disequilibrium processes are implicated in the formation of all the meteoritic material listed above. Chemical zoning is frequently observed in the crystals formed from melt in igneous objects. Some chondrules contain heterogeneities in alkaline element contents and silicate mineral zoning  that may have resulted from gas-melt interactions (continued condensation) during cooling (*Matsunami et al.*, 1993; *Tissandier et al.*, 2002; *Krot et al.*, 2004a,b; *Alexander and Grossman*, 2005). Matrix assemblages of amorphous + crystalline silicates have been attributed both to disequilibrium condensation (amorphous) followed by crystallization during annealing (e.g., *Brearley*, 1993), or to equilibrium condensation (crystalline) followed by amorphitization by radiation damage (*Zolensky et al.*, 2003). Fractional evaporation may result in PGE-rich nuggets (*Cameron and Fegley*, 1982), and is certainly important in forming some CAIs (*Davis*, 2003). Partial re-condensation has been invoked in several contexts, for example to explain compositions of rims on CAIs (*Wark and Boynton*, 2001).

### 4.3.  Refractory Al-Ca-Si-Mg Phases

*4.3.1. Ca-Aluminates.*  In most cosmochemically plausible protoplanetary chemical systems with C/O < 1, where liquid is not thermodynamically stable, corundum ($Al_2O_3$) is the highest-T solid to condense (Plates 1, 3a, and 4; Fig. 1). Nearly pure corundum, enclosed in hibonite, with perovskite or grossite, has been reported in five inclusions (*Bar-Matthews et al.*, 1982; *Fahey*, 1988; *Krot et al.*, 2001). At least one (*Simon et al.*, 2002) is convincingly not melted, nor formed by reactions among less refractory components, nor formed by evaporation (*Floss et al.*, 1998). The grossite, perovskite, and spinel ($MgAl_2O_4$) found with hibonite in the most refractory CAIs are nearly pure phases (Table 1; e.g., *Weber and Bischoff*, 1994; *Simon et al.*, 1994; *Brearley and Jones*, 1998; *Aléon et al.*, 2002). Ca-monoaluminate has been reported in exactly one occurrence, with grossite, perovskite and melilite (*Ivanova et al.*, 2002). These assemblages are consistent with predictions in Plate 1 for fractional condensation at high T and low P$^{tot}$, and at small dust enrichments (Plate 5).

Hibonite is calculated to condense after corundum in the absence of liquid in systems with C/O < 1. Hibonites in CAIs can contain up to 6 weight % TiO$_2$+Ti$_2$O$_3$ (*MacPherson and Grossman*, 1984; *Brearley and Jones*, 1998; *Aléon et al.*, 2002). The Ti$^{+3}$/Ti$^{+4}$ ratio in hibonite is a sensitive indicator of oxygen fugacity (*Ihinger and Stolper*, 1986; *Beckett et al.*, 1988). Spherules with nearly pure hibonite, surrounded by spinel, with or without perovskite, are rare components of several meteorite types which have been vigorously investigated (*MacPherson et al.*, 1983; *Hinton et al.*, 1988; *Ireland et al.*, 1991; *Floss et al.*, 1998; *Simon et al.*, 2002). Textures of some inclusions (e.g., SH-6 of *MacPherson et al.*, 1984) provide compelling evidence of vapor-solid condensation of euhedral hibonite plates, and subsequent reaction with vapor to directly form spinel. *Beckett and Stolper* (1994) argued, based on experimental results, that hibonites in the core regions of melted Allende type A CAIs are surviving precursor grains.

***4.3.2. Ca-, Al-Silicates.*** Melilite and Ca-rich pyroxene are, with spinel, the most abundant constituent minerals of CAIs that are found in all chondrite classes except CI (*Brearley and Jones*, 1998; *Fagan et al.*, 2000; *MacPherson*, 2003; *Beckett et al.*, 2005). They sometimes contain measurable Sc and Zr (*El Goresy et al.* 2002; *Simon et al.*, 1991, 1996), and are highly depleted in more volatile elements (Na, Fe). Zoning in melilite crystals can be diagnostic of their origin, whether they crystallized from a closed-system melt (igneous origin), a cooling melt in chemical equilibrium with cooling vapor (metasomatic igneous), or directly from cooling vapor either fractionally or in equilibrium (condensation *senso stricto*). Anorthite appears to have formed by a vapor-solid reaction in many CAIs.

The least melted CAIs record complex processes in the textural relationships among the mineral phases. These textures are disequilibrium phenomena that record sequential, incomplete and partial equilibration of minerals with surrounding vapor. *MacPherson* (2003) and *Beckett et al.* (2005) have reviewed Wark-Lovering rim mineralogy and reverse zoning of melilite in CAIs. An unresolved problem is the enclosure of spinel in melilite in most type A CAIs, because spinel is calculated to form at lower T than melilite in most conditions (Plates 1-4; Fig. 3). *Beckett and Stolper* (1994) argued that spinel growth on a hibonite substrate would be kinetically favored (over melilite growth) based on their crystallochemical similarities. Alternatively, *Beckett* (1986) suggested that removal of hibonite-bearing objects rich in Al and the most refractory REE would both suppress melilite condensation relative to spinel, and also explain Group II REE patterns.

In the condensation calculations shown here, the Al-, Ti-rich pyroxenes formerly known as 'fassaite' (now, subsilicic titanoan aluminian pyroxene) become stable at higher T than diopside (CaMgSi$_2$O$_6$), consistent with the experiments by *Stolper* (1982). The calculated compositions of those fassaites cannot be accurate, however, given the lack of Ti$^{+3}$ in the model pyroxene. Although Ti$^{+4}$ does not readily enter tetrahedral sites in silicates, the behavior of the reduced species Ti$^{+3}$ is relatively unknown, and could have significant effects on fassaite stability relative to other phases. *Yoneda and Grossman* (1995) modified the estimates of pyroxene thermodynamic properties by *Beckett* (1986), and assumed ideal solution, to simulate condensation of fassaite from solar and dust-enriched vapors at various total pressures (P$^{tot}$) below one bar. They condensed fassaite at slightly lower temperature than melilite, but their predicted compositions (their Fig. 9) correspond poorly to observed CAI fassaites (e.g., *Beckett*, 1986, Table 6; *Simon et al.*, 1998). Scandium may also stabilize pyroxenes at high T. The thermodynamic properties of refractory Ca-, Al-, Ti-, Sc-pyroxenes remain mysterious. Sequential, incomplete equilibration of CAI pyroxenes with surrounding vapor is indicated by their chemical zoning (*Brearley and Jones*, 1998), particularly in CAI rims. Texturally late FeO-enrichment probably results from aqueous alteration of rim pyroxenes on chondrite parent asteroids.

### 4.3.3. Rare Earth Elements.

Hibonite is the mineral stable at the highest temperatures that incorporates significant REE (*MacPherson and Davis*, 1994; *Ireland and Fegley*, 2000; *Simon et al.*, 2002; *Lodders*, 2003). REE enrichments of from 5 to 100, and 0.6 to 5 times CI chondritic abundances are seen, respectively, in CAIs (*Russell et al.*, 1998) and in rapidly quenched chondrules (*Engler et al.*, 2003) with unfractionated REE patterns. These objects may retain the full complement of refractory lithophile REEs condensed directly into their precursors. Other CAIs and chondrules have patterns depleted in the more refractory REE (group II pattern; cf., *Mason and Taylor*, 1982). These depletions have been interpreted to result from the earlier condensation and removal of an ultra-refractory component which retained primarily the more refractory REEs (*Boynton*, 1975, 1978; *Simon et al.*, 1994; *Ireland and Fegley*, 2000). Their volatility-related REE abundances, other isotopic constraints, and high melting temperatures, have been seen as evidence for CAI and chondrule origins as condensates near the sun or in the sun's photosphere, and subsequent transport outward by stellar winds (*Shu et al.*, 2001; *Ireland and Fegley*, 2000). This idea resonates with the reasoning, before any knowledge of REEs, of *Sorby* (1877) who argued that "some at least of the constituent particles of meteorites were originally detached glassy globules, like drops of fiery rain".

## 4.4. Metal Alloy and Ferromagnesian Silicates

### 4.4.1. Metal and Siderophile Elements.

Metallic iron-nickel alloy is found as sub-grains in presolar rocks (*Bernatowicz et al.*, 1999), and is common in chondritic meteorites. The chemical state of iron (metal, sulfide, oxide) in chondrites, and the relative abundance of iron among H, L, and LL ordinary chondrites, may result from fractionation of metals and S caused by their relative volatilities or susceptibility to oxidation and reduction in different regions of the protoplanetary disk. Efforts to trace these processes are focussed on the behavior of trace elements (siderophile) that dissolve in Fe (e.g., *Blum et al.*, 1989). Some of the platinum-group elements (PGE) are extremely refractory (Mo, Ru, Os, W, Re, Ir and Pt), others less so (Fe, Ni, Pd, Co, Rh, Au, Cu) (*Palme and Wlotzka*, 1976; *Humayun et al.*, 2002). The redox behavior of the siderophiles varies as well. Assemblages of PGE-rich grains in CAIs have chondritic (solar) abundances of metals, although individual nuggets are highly heterogeneous (*MacPherson*, 2003).

Zoned metal grains in some CB and CH chondrites have core-to-rim gradients in composition that are correlated to the relative volatilities of the metals. *Campbell et al* (2001) could not reconcile zoning patterns with equilibrium fractional condensation. Yet these zoned grains have been interpreted by others to form by sequential equilibration with vapor, in calculations that account for diffusive processes (*Meibom et al.*, 2001; *Petaev et al.*, 2003). These models are capable of quantitatively matching many features of the metal grains, but they require a different set of parameters for each grain (*Petaev et al.*, 2001). Exactly how these zoned metal grains formed is uncertain, pending better diffusion data on PGEs in metal.

Metal grains in chondrules have also been investigated for evidence of interaction with vapor. Recondensation is proposed to explain metal grains rimming CR chondrules, and enriched in more volatile siderophile elements: Initially FeS- and FeO-bearing chondrules are heated and partially evaporate; crystals coarsen; metal is reduced and forms alloy grains, while volatile enriched metal vapor recondenses as rims (*Kong and Palme*, 1999; *Zanda et al.*, 2002). Alternatively, reduction occurs by equilibration with either a reducing gas, or carbon present with the precursors (*Huang et al.*, 1996; *Hewins et al.*, 1997). Only a few wt% C are necessary to reduce most of the fayalite in typical chondrules (*Connolly et al.* 1994). *Connolly et al.* (2001) concluded from experimental work (cited therein) that vapor is a less likely reductant than included C (cf., *Hanon et al.*, 1998). *Lauretta et al.* (2001) came to a similar conclusion, and estimated 4.6 wt% Si in chondrule metal formed by this process in ordinary chondrites, far higher than would condense in an oxidizing (solar or dust enriched) environment (*Grossman et al.*, 1979; *Ebel and Grossman*, 2000). They went further, and assumed volatile (O, S, and Na) addition after metal formation, through interaction with surrounding vapor at lower T, and used commercially available code (HSC, *Eriksson and Hack*, 1990) to predict the corrosion of Si-, P-, Ni-bearing metal to silica ($SiO_2$), troilite (FeS), and other fine-grained phases they observed in SEM images. One difficulty with the *in situ* C-reduction hypothesis is the absence of C from metal in carbonaceous chondrites (*Mostefaoui et al.*, 2000).

   ***4.4.2. Mg-(Fe)-Silicates.*** The elements Fe, Mg, and Si are ~10 times more abundant than the more refractory Al and Ca. Olivine (Mg-orthosilicate, Table 1) has been called 'astrophysical silicate', because the olivine species forsterite ($Mg_2SiO_4$) is the most refractory silicate to condense from a solar gas in quantity. For example, at $P^{tot}=10^{-3}$ bar, ~0.0027% of all atoms are condensed at 1444 K when forsterite appears, and ~0.0150% are condensed as olivine, out of ~0.0165% total, at 1370 K. This small T (or, alternatively, $P^{tot}$) interval coincides with the transition from refractory inclusion (CAI) assemblages to chondrule mineral assemblages. It is unlikely that most meteoritic inclusions or their precursors passed through this interval in a smooth, monotonic way; they probably experienced fluctuations in T and $P^{tot}$.

   Olivine is the most abundant silicate in most chondrites, however there is only controversial evidence that any olivine is a primary condensate, or reached equilibrium with vapor (*McSween*, 1977; *Olsen and Grossman*, 1978; *Palme and Fegley*, 1990; *Steele*, 1986, 1995; *Weisberg et al.*, 1997; *Weisberg and Prinz*, 1996, 1998; *Weinbruch et al.*, 2000; *Pack et al.* 2004). The FeO and $Na_2O$ content of many meteoritic silicates has been altered during secondary processing on parent bodies, obscuring the signatures of pre-accretionary processes (*Bischoff*, 1998; *Sears and Akridge*, 1998; *Krot et al.*, 2000; *Brearley* 2003). If all Fe in a gas of solar composition (Mg/Si ~1.07) existed as metal, orthopyroxene (enstatite, $MgSiO_3$) would dominate the silicate assemblage, as it does in enstatite chondrites. In condensation from solar gas, FeO-bearing silicates only form by reaction of metal and metal sulfide with Mg-silicates and vapor at very low temperatures, where diffusion is slow. Alternatively, FeO-bearing chondrules may have formed from flash-melted FeO-bearing precursors that crystallized at low temperatures or in an oxidizing environment, without losing FeO by evaporation during melting. Negligible heavy Fe isotopic enrichment is observed in chondrule silicates (*Zhu et al.*, 2001; *Alexander and Wang*, 2001; *Mullane et al.*, 2003). Flash-melting would then have had to occur in a vapor enriched in Fe (*Galy et al.*, 2000; *Alexander*, 2001; *Ebel*, 2005). Similar issues attend the Na content of chondrules and CAIs (e.g., *Sears et al.*, 1996).

   One way to increase the FeO content of silicates when they first become chemically stable is for the surrounding vapor to be enriched in oxygen. *Ebel and Grossman* (2000, their Fig. 8) showed that above 1200 K, systems enriched in chondritic vapor yield ferromagnesian silicates with molar FeO/(FeO+MgO) ratios of 0.1 to 0.4, and also stabilize chondrule liquids against evaporation (Plate 4). Alternatively, enrichment of vapor in $H_2O$ ice can stabilize FeO in silicates at high temperature (*Krot et al.*, 2000, their Fig. 6). Mechanisms capable of melting chondrule precursors (shock waves, *Desch and Connolly*, 2002; current sheets, *Joung et al.*, 2004) would simultaneously vaporize small silicate dust grains, providing a local dust-enriched region during chondrule cooling.

## 4.5.  Carbides, Sulfides, and Phosphides

Pre-solar grains have survived ejection from AGB stars or novae, interstellar processing (*Bernatowicz*, 2005; *Nuth et al.*, 2005), presolar cloud collapse (*Boss and Goswami*, 2005) heating and irradiation in the early solar nebula (*Chick and Cassen*, 1997; *Connolly et al.*, 2005; *Chaussidon and Gounelle*, 2005), accretion into chondrite parent bodies (*Cuzzi and Weidenschilling*, 2005; *Weidenschilling and Cuzzi*, 2005), followed by aqueous alteration (*Brearley*, 2005), impacts (*Bischoff et al.*, 2005), lithification, and thermal metamorphism on the parent bodies (*Huss and Lewis*, 1995; *Zinner*, 1998; *Mendybaev et al.*, 2002; *Krot et al.*, 2005; *Huss et al.*, 2005). Graphite, SiC, TiC, Fe-Ni alloy, spinel, corundum and other presolar phases are all refractory minerals predicted to condense from vapors with variable C/O ratio but solar proportions of other elements (Plate 3b; *Ebel*, 2000). Isotopic constraints on grains formed in supernovae require mixing between different chemical zones of supernovae, if the grains condensed in systems approaching equilibrium (*Travaglio et al.*, 1999; *Ebel and Grossman*, 2001; *Meyer and Zinner*, 2005). Quantitatively addressing the condensation of these rocky materials would appear to require better models of zone mixing in supernova explosions, to reduce the universe of possible mixtures capable of producing the observed condensates.

Enstatite chondrites contain mixed phases in the FeS-MnS-MgS system (niningerite, alabandite), which may record primary equilibrium with vapor (*El Goresy et al.*, 1988; *Crozaz and Lundberg*, 1995; reviewed by *Brearley and Jones*, 1998). Metal (Plate 3b) is Si-rich, and the Si content of metal increases steadily as coexisting vapor becomes more highly reducing, until the stability fields of Fe-Si intermetallic compounds are reached at high C/O ratios of the bulk system. These compounds are not observed in meteorites. Other common phases include oldhamite, nearly pure CaS; troilite, FeS; daubreelite, (Fe, Mn, Zn)$Cr_2S_4$; schreibersite, $FeNi_3P$; and perryite, $\sim(Ni,Fe)_2(Si,P)$. Other sulfides have been discovered in enstatite chondrites, but their cosmochemical significance is unknown (e.g., *Nagahara*, 1991). The role of vapor-solid equilibria in the origin of components of enstatite chondrites remains poorly constrained.

## 5. DISCUSSION AND CONCLUSIONS

## 5.1.  The 50% Condensation Temperature

*Wasson* (1985, his Table G1) has compiled a table of "50% condensation temperatures" for the elements from previous calculations (e.g., *Wai and Wasson*, 1979). *Lodders*, 2003). This single metric has been useful in categorizing the elements by their volatility in a gas of solar composition at fixed $P^{tot}$ (e.g., *Hobbs et al.*, 1993), but can be misleading because elements condense over different T ranges (*Ebel*, 2000). *Lodders* (2003) treated the entire periodic table in a self-consistent manner, to calculate 50% condensation temperatures at $P^{tot} = 10^{-4}$ bar, using her assessment of the most recent elemental abundance data for a vapor of solar composition (see Section 3.1.2).

## 5.2.  Contexts of Condensation and Gas-Vapor Equilibrium

Incontrovertible evidence for direct condensation of rocky meteoritic material from vapor to solid or liquid phases remains elusive, apart from micron- to submicron-sized presolar grains formed around other stars, probably in single outflow episodes. These grain sizes cannot be taken as exemplary of solar nebular processes. Except for some isotopic anomalies that may have resulted from local processes (e.g., O self-shielding, *Navon and Wasserburg*, 1985; *Clayton*, 2004), the initial gas and dust of the protoplanetary disk appear to have been thoroughly mixed (*Palme*, 2001). Most grains that may represent direct vapor-solid condensates, even in unmelted IDPs, are isotopically solar at the spatial resolution of current instruments. It is not clear how this homogenization occurred, if the solar system was hot only inside ~3 AU, unless mixing in the accretion shock was highly efficient. Perhaps the interstellar medium, ultimate source of precursor solids, is also mostly well-mixed, most grains are sub-micron sized, and a 'scoop' of them is essentially solar (*Brearley*, 1993). It is also possible that thin planar heating events, such as current sheets and shock waves, driven by magnetorotational and gravitational instabilities, were commonplace in different regions of the protoplanetary disk well beyond 3 AU. These forces would have driven repeated cycling of micron-sized solids through the vapor phase, building up a supply of ever-larger condensate grains, and depleting regions of the disk in the same volatility-controlled patterns observed in dense molecular clouds (*Ebel*, 2000). Resulting larger dust and dust aggregates would be isotopically close to solar if evaporating materials were unmelted. Both shocks and current sheets would produce local T excursions decreasing with radial distance from the central star. This is one context for direct condensation in the early solar system. Future astronomical observations of disks at high spatial resolution, by coronography and/or interferometry, should reveal whether such fractionation exists and at what radii from central stars. The precursors to chondrules and CAIs were probably similar to IDP components, isotopically solar in most respects, and fractionated into Ca-, Al-, Ti-rich oxides; Mg-, Si-rich silicates; metal; and weakly bonded components rich in volatile elements.

*Rietmeijer* (1998) has developed a theory of 'hierarchical accretion' in explaining the textures and compositions of IDPs. Meteoritic inclusions experienced many episodes of heating to high temperatures, so their textures would not reveal hierarchical accretion as clearly as do IDPs. But the paradigm may apply most evidently to more refractory CAIs, and to AOAs, where sequential layers of minerals, mechanically accreted, were also subsequently changed by sequential metasomatic reactions with vapor. Many objects preserve layers of minerals of different volatility, as if they passed through regions rich in, and accreted rims of, dust that had been condensed at a particular temperature, $P^{tot}$, and vapor composition (*MacPherson et al.*, 1985). As the objects themselves occasionally encountered more than one such heating event, dust would progressively sinter and acquire an igneous texture, becoming part of the chondrule or CAI, and also reacting with gas, for example a gas with higher $P_{SiO}$ than encountered previously. Hierarchical accretion may be caught in the act, in concentrically zoned CAIs (e.g., *Simon et al.*, 1994), or in chondrules with thick layers of dusty rim material, and with concentric internal structure. Partial or complete remelting of such objects would sinter the rim or homogenize the entire object, even as it accreted still more dusty material. This scenario requires a high dust to gas ratio, with heating events evaporating nearby fine dust particles so that vapor pressures of oxygen and condensable elements are locally increased around melted chondrules. This would suppress evaporative enrichment in heavy isotopes, make liquids stable against evaporation, and increase the stable FeO content of silicates. This kind of accretion, resulting in chondrules and CAIs, would occur in disk regions closer to the sun, with higher densities of dusty material.

Vapor-solid-liquid equilibria appear to have been approached inefficiently, due to the rapid time scales of most heating events, resulting in the abundant evidence of disequilibrium in the heterogeneous and zoned mineral grains of so many meteoritic inclusions. It is also likely that kinetic and surface effects are important in stabilizing particular minerals relative to others, in direct gas-solid reactions (*Beckett*, 1986). Differences in thermochemical stability between these phases are small, so surface energy effects become important at small grain sizes. Surface forces may affect macroscopic grain compositions, as they do in sector-zoned crystals. Displacive reactions of preexisting, more refractory phases (e.g., melilite) with vapor during cooling might also be energetically favored, relative to reconstructive reactions (i.e., adjusting a few *versus* breaking many bonds). This may explain textures in unmelted objects, in which anorthite appears to have formed at a higher T than olivine (e.g., *Krot et al.*, 2002; *Weisberg et al.*, 2005).

# 6. OUTLOOK

***6.1. Observations.*** The equilibrium chemical behaviors of the abundant elements (Si, Fe, Ni, Mg, Al, Ca, Ti) are well understood except in highly reduced systems. Attention has shifted to the less abundant elements. We can expect increasing application of new, high-resolution analytical techniques to elemental and isotopic abundances in particular components of meteorites, and correlation of these results with other evidence (e.g., *Mullane et al.*, 2001; *Young et al.*, 2002; *Kehm et al.*, 2003; *Friedrich et al.*, 2003; *Dauphas et al.*, 2004; *Johnston et al.*, 2004). New techniques of laser ablation ion-coupled plasma mass spectrometry (LA-ICPMS) enable investigation of ever-smaller spatial regions of metal grains (e.g., 30micron x 15micron deep) at very high mass resolution (*Campbell and Humayun*, 1999). The moderately volatile elements (e.g., Mn, Cu, Na, Se, Zn, Cd) are increasingly accessible to chemical and isotopic analysis in meteoritic materials. High compositional accuracy and fine spatial resolution of analyses of these and other trace elements will allow new constraints to be placed on the vapor-solid-liquid equilibria operative in the formation of chondritic components.

The search for extrasolar planets is driving astrophysical observations of *disks* around nearby young stars that will yield images of increasingly higher spatial and spectral resolution. It will become possible to ascribe radial gradients in elemental abundances of disk gas to grain formation and/or destruction. Better constraints on temperatures in evolving disks should also result from these observations. It is hoped that these results will have direct relevance to the origin of the rocky material in chondritic meteorites.

***6.2. Theory.*** Condensation calculations relevant to meteorites are grounded in over a century of systematic experimental petrology, mineralogy, and basic research (e.g., *Bowen*, 1914; *Ferguson and Buddington*, 1920; *Robie et al.*, 1978; *Yoder*, 1979; *Chase*, 1998). Results of this difficult work, with modern computational tools, provide the means to calibrate mathematical models for the thermodynamic behavior of minerals and melts (*Charmichael and Eugster*, 1987; *Geiger*, 2001). These fields are perhaps not as actively supported as they once were. The public availability of high-quality primary experimental data by metallurgists and materials scientists appears to have declined since the 1970s. At the same time, the computer has enabled collation and evaluation of subsets of the mountain of existing data, resulting in open databases (e.g., *Berman*, 1988; *Kuzmenko et al.*, 1997) usually optimized for particular subsets of the cosmochemical universe (e.g., THERMOCALC for metamorphic petrology, by *Holland and*

*Powell*, 1998;). Proprietary databases (e.g., FactSage, *Bale et al.*, 2002) also exist, but their reliability is difficult to assess.

Strong backgrounds in thermodynamics and experimental petrology are necessary for critical assessment of existing data; for the building of optimized, internally consistent databases, both for equilibrium and kinetic approaches; for improvement of solid and liquid solution models applicable to extraterrestrial rocky materials; and for informed use of computational tools. Non-open source, 'black-box' software is increasingly available to perform equilibrium thermodynamic calculations with some degree of presumed rigor. These must be used with caution. Progress is most likely to result from the combination of kinetic constraints with thermodynamic models, in open-sourced codes. Early and rigorous training of students in thermodynamics, isotope chemistry, and experimental petrology must be the foundation for future progress in reading the solar system's earliest rocks.

**6.3. Experiment.** New techniques are being developed to directly probe condensation phenomena (e.g., *Toppani and Libourel*, 2003; *Toppani et al.*, 2005). The Knudsen cell has, belatedly, begun to see more application in petrologically relevant experiments (e.g., *Dohmen et al.*, 1998). The combination of laser heating devices and quadrupole mass spectrometers to measure gas speciation promises to allow direct investigation of both condensation and evaporation. The parameters necessary to model evaporative mass fractionation are being discovered in the lab (*Davis et al.*, 1990; *Richter et al.*, 2002; *Dauphas et al.*, 2004). Amorphous grains 20-30nm in size condense via homogeneous nucleation from hot gases quenched in short times (*Nuth et al.*, 2002). Their annealing to crystalline phases as a function of time and temperature has been quantified with application to observed AGB outflows, and to make inferences about protoplanetary disk grain cycling (*Brucato et al.*, 1999; *Nuth et al.*, 2002). Other recent work at slightly longer timescales has resulted in partially or fully crystalline olivine condensed from vapor (*Tsukamoto et al.*, 2001). Even longer timescales are being explored, in heterogeneously nucleated systems (*Toppani et al.*, 2004). Gas-grain reactions at lower temperatures (< 600 K) are also being explored (*Llorca and Casanova*, 1998). These kinds of experiments become most relevant to the processes recorded by objects in meteorites, when combined with kinetic models grounded in an understanding of equilibrium phase relations. We can expect substantial progress both in direct laboratory explorations, and their understanding using better models, by the next edition of this volume.

**Acknowledgments**. This research has made use of NASA's Astrophysics Data System Bibliographic Services. This work was partially supported by the NASA Cosmochemistry program under grant NAG5-12855.